\documentclass[aps,pra,twocolumn,amsmath,amssymb,superscriptaddress]{revtex4-2}

\usepackage{graphicx}
\usepackage{color}
\usepackage[hidelinks]{hyperref}

\hypersetup{colorlinks=true, linkcolor=black, citecolor=black, urlcolor=blue}

\begin{document}
\author{Karol Gietka}
\email[]{karol.gietka@uibk.ac.at}

\affiliation{Institut f\"ur Theoretische Physik, Universit\"at Innsbruck, A-6020 Innsbruck, Austria} 
\affiliation{Quantum Systems Unit, Okinawa Institute of Science and Technology Graduate University, Onna, Okinawa 904-0495, Japan} 
\author{Helmut Ritsch}
\affiliation{Institut f\"ur Theoretische Physik, Universit\"at Innsbruck, A-6020 Innsbruck, Austria} 

\title{Squeezing and overcoming the Heisenberg scaling \\ with spin-orbit coupled quantum gases}

%%%%%%%%%%%%%%%%%%%%%%%%%%%%%%%%%%%%%%%%%%%%%%%%%%%%%%%%%%%%%%%%%%%%%%%%%%%%
%%%%%%%%%%%%%%%%%%%%%%%%%%%%%%%  ABSTRACT  %%%%%%%%%%%%%%%%%%%%%%%%%%%%%%%%%
%%%%%%%%%%%%%%%%%%%%%%%%%%%%%%%%%%%%%%%%%%%%%%%%%%%%%%%%%%%%%%%%%%%%%%%%%%%%

\begin{abstract}
We predict that exploiting spin-orbit coupling in a harmonically trapped spinor quantum gas can lead to scaling of the optimal measurement precision beyond the Heisenberg scaling. We show that quadratic scaling with the number of atoms can be facilitated via squeezed center-of-mass excitations of the atomic motion using a 1D spin-orbit coupled fermions or strongly interacting bosons (Tonks-Girardeau gas). Based on predictions derived from analytic calculations of the corresponding quantum Fisher information, we then introduce a protocol, which overcomes the Heisenberg scaling (and limit) with help of a tailored excited and entangled many-body state of a non-interacting Bose-Einstein condensate. We identify corresponding optimal measurements and argue that even finite temperature as a source of decoherence is, in principle, rather favorable for the obtainable precision scaling.
\end{abstract}

\date{\today}
\maketitle

%%%%%%%%%%%%%%%%%%%%%%%%%%%%%%%%%%%%%%%%%%%%%%%%%%%%%%%%%%%%%%%%%%%%%%%%%%%%
%%%%%%%%%%%%%%%%%%%%%%%%%%%%%%%  INTRODUCTION  %%%%%%%%%%%%%%%%%%%%%%%%%%%%%
%%%%%%%%%%%%%%%%%%%%%%%%%%%%%%%%%%%%%%%%%%%%%%%%%%%%%%%%%%%%%%%%%%%%%%%%%%%%

\section{Introduction}
Quantum metrology aims at exploiting quantum resources to increase the precision of measurements~\cite{Giovanetti2006QuantumMetrology,Giovanetti2011advancesinQM}. One of the most widely used techniques in precision metrology is Ramsey interferometry~\cite{pezze2018QM_RMP}. There,  metrology theory for the measurement of the atomic transition frequency $\Omega$ using an ensemble of $N$ independent two-level atoms predicts an inverse square root scaling of the relative precision with respect to the number of atoms $N$. This follows from the law of large numbers which in this case corresponds to a large number of atoms. In order to improve the $1/\sqrt{N}$-scaling, one can harness quantum correlations, for example, entanglement. Using the quantum spin squeezing~\cite{wineland1992squeezing,MA2011quantumSS} can elevate the scaling to $\propto 1/N^{5/6}$;  ultimately, exploiting even stronger entangled states allows to achieve the so-called Heisenberg scaling $\propto 1/N$. In the optimal case, the numerical factor in front of the scaling is equal to 1 which then gives a concrete limit for any $N$. To be more precise one finds the so-called standard quantum limit for separable atoms and the Heisenberg limit for a maximally entangled state~\cite{2012rafaljan}.

It seems then that the only way to further increase the precision is to increase the number of atoms $N$. From a quantum-mechanical theory construction viewpoint, increasing the number of atoms of a given system amounts to increasing the size of the Hilbert space. From this perspective, it thus becomes apparent that the size of the Hilbert space is related to the optimal precision obtainable in a measurement. Building on this rather basic and simple observation we will show, how coupling a harmonic oscillator degree of freedom to the two-level atoms can lead to multiplicative enhancement of the Heisenberg scaling or even arrive at a more favorable scaling. To this end, we will consider the generic case of a one-dimensional system of ultra-cold quantum atoms, fermions, or interacting bosons, where the harmonic oscillator simply represents the motional degree of freedom of the atoms, which are anyway typically trapped in a harmonic potential. The protocols that we consider rely on an adiabatic preparation of the ground state and therefore are only weakly affected by decoherence and loss of excitations. As we will outline later, for a thermal state at finite temperatures, where the atoms occupy excited states, the optimal possible measurement precision counterintuitively might be even further increased as such states exhibit an increased number of excitations. Following this observation, we identify an excited state of a non-interacting Bose-Einstein condensate that can also exhibit super-Heisenberg scaling of the measurement precision.% without resorting to entanglement.

%%%%%%%%%%%%%%%%%%%%%%%%%%%%%%%%%%%%%%%%%%%%%%%%%%%%%%%%%%%%%%%%%%%%%%%%%%%%
%%%%%%%%%%%%%%%%%%%%%%%%%%%  SPIN-ORBIT COUPLING  %%%%%%%%%%%%%%%%%%%%%%%%%%
%%%%%%%%%%%%%%%%%%%%%%%%%%%%%%%%%%%%%%%%%%%%%%%%%%%%%%%%%%%%%%%%%%%%%%%%%%%%

\section{spin-orbit coupling}
The essential ingredient that we introduce to go beyond standard scaling is the coupling between some (typically two) internal atomic states and the motional harmonic oscillator representing the trap. This facilitates \emph{transfer} of the information about the atomic transition frequency to the harmonic oscillator. Therefore increasing the size of the (parameter-dependent) Hilbert space and thus the capacity to represent information about the unknown parameter in the collective atomic state. As a simple example, we consider here basic spin-orbit coupling~\cite{2016propertiessocbec} which can be directly experimentally implemented~\cite{2011socbec,2016socbecketterle}. It relies on coupling the two atomic levels differently to their momenta. For a non-interacting gas of bosons in a one-dimensional harmonic trap, each atom is described by a single-particle Hamiltonian (we set $\hbar=1$ throughout the entire manuscript)
\begin{align}\label{eq:socH}
    \hat H = \frac{ \left(\hat p + k \hat \sigma_z\right)^2}{2m} + \frac{m \omega^2 \hat x^2}{2} + \frac{ \Omega}{2}  \hat \sigma_x
\end{align}
where $\hat x$ and $\hat p \equiv - i \partial_x$ are position and momentum operators of an atom, $\omega$ is the frequency of the harmonic trap, $m$ is the atomic mass, $\Omega$ is the atomic frequency, $\hat \sigma_i$ are the standard Pauli matrices, and $k$ is the spin-orbit interaction strength. Depending on the coupling strength $k$, one can distinguish two phases. The single minimum (normal) phase is characterized by $\sqrt{m \Omega/2} > k$, and a double minimum (stripe) phase is characterized by $ \sqrt{m\Omega/2} < k$. In this work, we only focus on the normal phase of the system; however, an extension to the (symmetry-broken) stripe phase is straightforward~\cite{2015qpt_qrm,2020CriticalParis}.

In the following, we consider a one-dimensional gas of $N$ two-level atoms in a harmonic trap along direction $x$. We assume that the ground state of the gas is spin-polarized which amounts to a condition of $N \omega \ll \Omega$. This guarantees that only the excited states of the harmonic oscillator will be occupied. To draw a parallel with light-matter systems we switch to the energy eigenbasis of the harmonic oscillator. By introducing annihilation and creation operators of the harmonic oscillator and applying the Schrieffer-Wolff transformation under the condition $\omega/\Omega \rightarrow 0$~\cite{2015qpt_qrm} (consistent with the polarized gas condition), we can rewrite the Hamiltonian as~(see Appendix~\ref{appendix:A} for details)
\begin{align}\label{eq:effectiveH}
     \hat H \approx \omega \hat a^\dagger \hat a + \frac{ \Omega}{2} \hat \sigma_z + k^2{\frac{  \omega}{2m \Omega}} (\hat a +  \hat a^\dagger)^2 \hat \sigma_z.
\end{align}
In this form, the two atomic energy levels are effectively decoupled and the $n$th eigenstate of the system is
\begin{align}
    |\psi_n(\xi) \rangle = \hat S(\xi) |n\rangle \otimes|\!\downarrow\,\rangle
\end{align}
with $\hat S(\xi) \equiv \exp\{(\xi/2)(\hat a^\dagger)^2-(\xi^*/2)\hat a^2\}$ being the squeeze operator where $\xi = -\frac{1}{4} \ln\{1-(k/k_c)^2\}$ and $k_c = \sqrt{\Omega \omega}$. These single-particle states will serve us now to build the many-body states, and calculate the quantum Fisher information~\cite{1994distancestates} which is a central object in quantum metrology~\cite{2009pezzesmerzientHL} as it is related to the precision of the measurement through the Cram\'er-Rao bound $\Delta \Omega \geq 1/\sqrt{\mathcal{I}_\Omega}$, where $\mathcal{I}_\Omega$ is the quantum Fisher information associated with (the unknown) parameter $\Omega$.

%%%%%%%%%%%%%%%%%%%%%%%%%%%%%%%%%%%%%%%%%%%%%%%%%%%%%%%%%%%%%%%%%%%%%%%%%%%%
%%%%%%%%%%%%%%%%%%%%%%%%%%%%%%%% FERMI GAS  %%%%%%%%%%%%%%%%%%%%%%%%%%%%%%%%
%%%%%%%%%%%%%%%%%%%%%%%%%%%%%%%%%%%%%%%%%%%%%%%%%%%%%%%%%%%%%%%%%%%%%%%%%%%%

\section{spin-orbit coupled quantum gases}
The fermionic anti-symmetric ground-state wavefunction of the $N$ particle system is constructed according to the Slater determinant
\begin{align}\label{eq:slaterdet}
  \Psi_\mathrm{F}(x_1,\ldots,x_N) \propto \mathrm{det}[\phi_i(x_j)], \quad i,j =1,2,\ldots,N,
\end{align}
where $\phi_i(x_j)$ is the $i$th eigenstate of the single-particle Hamiltonian of the $j$th particle. We assume now that the unknown parameter is $\Omega$ (similar calculations can be performed treating other parameters of the system as unknown). The quantum Fisher information can be calculated using the derivative of the wavefunction with respect to the unknown parameter according to $\mathcal{I}_\Omega =\langle \partial_\Omega \psi|\partial_\Omega \psi\rangle - \langle \psi|\partial_\Omega \psi \rangle^2$~\cite{1994distancestates}. Upon acting with the derivative operator on the many-body wavefunction, it can be further simplified to $\mathcal{I}_\Omega = 4\Delta^2\hat{\mathcal{H}} \equiv 4\left(\langle \hat{\mathcal{H}}^2\rangle -\langle \hat{\mathcal{H}}\rangle^2 \right)$, with $\hat{\mathcal{H}} = \sum_{j=1}^N \hat h_j$ where 
\begin{align}
    \hat h_j = \frac{k^2}{k_c^2}\frac{i}{\frac{k^2}{k_c^2}-1}\frac{1}{8\Omega}\left( \hat a_j^\dagger \hat a_j^\dagger - \hat a_j\hat a_j\right)
\end{align}
is an operator acting on the $j$th particle.
It can be easily shown that $\langle \Psi(\xi)|\hat{\mathcal{H}}|\Psi(\xi)\rangle=0$, therefore we need to calculate only the second moment of $\hat{\mathcal{H}}$. We divide now this calculation into two steps (see Appendix~\ref{appendix:B} for details). First, we calculate the terms containing $\hat h_j^2$ of which all are equal. A straightforward calculation yields
\begin{align}
\begin{split}
    \langle \Psi(\xi)| \hat h_j^2 |\Psi(\xi)\rangle = \sum_{n=0}^{N-1} \frac{(1+n+n^2)}{64\Omega^2 N}\frac{k^4}{k_c^4}\frac{1}{\left(1-\frac{k^2}{k_c^2}\right)^2}.
\end{split}
\end{align}
Second, we calculate terms containing tensor products between $\hat h_j$ and $\hat h_k$ of which all are also equal. A straightforward calculation yields
\begin{align}
    \begin{split}
        \langle \Psi(\xi)| \hat h_j\hat h_k |\Psi(\xi)\rangle = -\sum_{n=0}^{N-1} \frac{n(n-1)}{8\Omega^2 N}\frac{k^4}{k_c^4}\frac{1}{\left(1-\frac{k^2}{k_c^2}\right)^2}.
    \end{split}
\end{align}
After collecting the terms together and evaluating the sums, the first contribution becomes
\begin{align}
    \frac{\frac{k^4}{k_c^4} N \left(N^2+2\right)}{24 \Omega ^2 \left(1-\frac{k^2}{k_c^2} \right)^2},
\end{align}
and the second one becomes
\begin{align}
 -\frac{\frac{k^4}{k_c^4} (N-2) (N-1) N}{24 \Omega ^2 \left(1 -\frac{k^2}{k_c^2}\right)^2}.
\end{align}
Interestingly, these two contributions include terms proportional to $N^3$; however, upon adding them to each other the $N^3$ terms cancel, and the quantum Fisher information becomes
\begin{align}
    \mathcal{I}_\Omega = \frac{\frac{k^4}{k_c^4} N^2}{8 \Omega ^2 \left(1-\frac{k^2}{k_c^2} \right)^2},
\end{align}
which exhibits Heisenberg scaling with respect to the number of atoms with additional improvement deriving from squeezing the center-of-mass excitations. This form of the quantum Fisher information assumes that the final state can be prepared instantaneously without consuming any time. In order to give it a metrological interpretation, we now assume that the coupling strength was adiabatically ramped from  $k/k_c =0$ towards $k_f/k_c<1$ (encoding thus information about $\Omega$) such that we always stay in the ground state manifold. The time of such an adiabatic sweep can be calculated to
\begin{align}
 T\approx\left(2\gamma\omega \sqrt{1-{k_f^2}/{k_c^2}}\right)^{-1}
\end{align}
where $\gamma\ll1$ is required to satisfy the adiabatic condition~\cite{2022criticalspeedup}. Inserting this time into the expression for the quantum Fisher information yields
\begin{align}\label{eq:qfifermion}
     \mathcal{I}_\Omega =\frac{\gamma^2 \omega^2 \frac{k_f^4}{k_c^4} N^2 T^2}{2 \Omega ^2 \left(1-\frac{k_f^2}{k_c^2} \right)},
\end{align}
where the $1-{k_f^2}/{k_c^2}$ in the denominator is related to squeezing of the center-of-mass excitations. This can be explicitly seen if we consider a single particle ground state for which the number of center-of-mass excitations becomes 
\begin{align}
    \langle \hat n\rangle= \sinh^2\xi \approx \left(4\sqrt{1-{k_f^2}/{k_c^2}}\right)^{-1}.
\end{align}

An interesting possibility would be the preparation of a bosonic state imitating the fermionic state from equation ~\eqref{eq:slaterdet}. This would result in quantum Fisher information exhibiting the super-Heisenberg scaling~(see Appendix~\ref{appendix:C} for details)
\begin{align}
   \mathcal{I}_\Omega  \xrightarrow{N\gg1} \frac{\gamma^2 \omega^2\frac{k^4}{k_c^4} N^3T^2}{6 \Omega ^2 \left(1-\frac{k_f^2}{k_c^2} \right)}.
\end{align}
%without using entanglement~\cite{2018noentrmp}.
Such a state, however, is not the groundstate of the system and thus its preparation might constitute a hindrance. Nevertheless, a bosonic ground state imitating the fermionic ground state wavefunction can be obtained by exploiting strongly interacting bosonic atoms.

%%%%%%%%%%%%%%%%%%%%%%%%%%%%%%%%%%%%%%%%%%%%%%%%%%%%%%%%%%%%%%%%%%%%%%%%%%%%
%%%%%%%%%%%%%%%%%%%%%%%%%%%%%%%% TONKS GAS  %%%%%%%%%%%%%%%%%%%%%%%%%%%%%%%%
%%%%%%%%%%%%%%%%%%%%%%%%%%%%%%%%%%%%%%%%%%%%%%%%%%%%%%%%%%%%%%%%%%%%%%%%%%%%

The so-called Tonks-Girardeau gas is a one-dimensional gas of strongly interacting bosonic atoms~\cite{1936tonks,1960girardeau,2004TGgas}. It possesses many fermion-like properties and is often referred to as \emph{fermionized} gas of bosons. This is reflected in the construction of the Tonks-Girardeau wavefunction which can be built from the fermionic wavefunction according to
\begin{align}
    \Psi_{\mathrm{TG}}(x_1,x_2,\ldots,x_N) = |\Psi_{\mathrm{F}}(x_1,x_2,\ldots,x_N)|,
\end{align}
where the TG and F subscripts indicate Tonks-Girardeau and fermionic wavefunction, respectively. Unfortunately, expressing it in terms of harmonic oscillator energy eigenstates does not lead to a \emph{bosonized} version of the state from equation~\eqref{eq:slaterdet}, and one should not expect cubic scaling of the quantum Fisher information with respect to the number of atoms. By inspecting the properties of the quantum Fisher information and the Tonks-Girardeau wavefunction, it is straightforward to show that the Tonks-Girardeau quantum Fisher information is equal to that of the fermionic system, which highlights yet another similarity between these two systems (see Appendix~\ref{appendix:C} for details).

%%%%%%%%%%%%%%%%%%%%%%%%%%%%%%%%%%%%%%%%%%%%%%%%%%%%%%%%%%%%%%%%%%%%%%%%%%%%
%%%%%%%%%%%%%%%%%%%%%%%%%%%%%%% comparison  %%%%%%%%%%%%%%%%%%%%%%%%%%%%%%%%
%%%%%%%%%%%%%%%%%%%%%%%%%%%%%%%%%%%%%%%%%%%%%%%%%%%%%%%%%%%%%%%%%%%%%%%%%%%%

\section{Comparison with the standard quantum limit and the Heisenberg limit}
Although the quantum Fisher information calculated in this work exhibits the Heisenberg or even super Heisenberg scaling, it is not clear yet when the considered strategy becomes better than the standard quantum limit of the frequency estimation and how it relates to the Heisenberg limit. The standard quantum limit of frequency estimation is $\Delta \Omega = {(\sqrt{N}T})^{-1}$. In order to overcome it, the following condition has to be satisfied
\begin{align}\label{eq:comparison}
    \frac{\mathcal{I}_\Omega}{NT^2} > 1 \rightarrow \frac{\gamma^2 \omega^2 \frac{kf^4}{k_c^2}N}{2 \Omega^2(1-\frac{k_f^2}{k_c^2})} \approx \frac{\gamma^2 \omega^2 N}{2 \Omega^2\left(1-\frac{k_f^2}{k_c^2}\right)} >1.
\end{align}
The above formula assumes that the center-of-mass excitations can be squeezed indefinitely. However, we expect that for every value of $\omega/\Omega$ there is a maximal number of squeezed excitations that can be generated. According to reference~\cite{2022commetrology} this can be upper bounded to 
\begin{align}
\langle \hat n \rangle  \approx \left(4\sqrt{1-{k_f^2}/{k_c^2}}\right)^{-1} < (\Omega/\omega)^{1/3}.
\end{align}
Plugging this into equation~\eqref{eq:comparison} gives the following condition for breaking the standard quantum limit of frequency estimation (up to numerical factors) 
\begin{align}
     N  >\frac{\Omega^{4/3}}{\gamma^2 \omega^{4/3}},
\end{align}
which is also the condition for the excited state of a non-interacting Bose-Einstein condensate to overcome the Heisenberg limit $\Delta \Omega = ({N T})^{-1}$.

%%%%%%%%%%%%%%%%%%%%%%%%%%%%%%%%%%%%%%%%%%%%%%%%%%%%%%%%%%%%%%%%%%%%%%%%%%%%
%%%%%%%%%%%%%%%%%%%%%%%%%%%%%%% MEASUREMENT  %%%%%%%%%%%%%%%%%%%%%%%%%%%%%%%
%%%%%%%%%%%%%%%%%%%%%%%%%%%%%%%%%%%%%%%%%%%%%%%%%%%%%%%%%%%%%%%%%%%%%%%%%%%%

\section{Optimal measurement}
\begin{figure}
    \centering
    \includegraphics[width=0.5\textwidth]{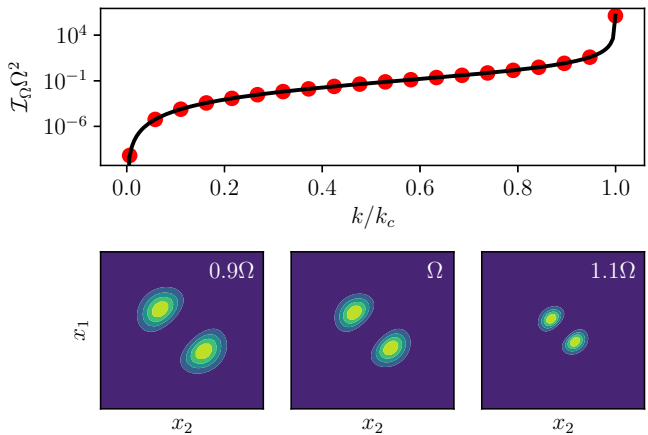}
    \caption{The top row shows the quantum (solid-black line) and classical (red dots) Fisher information as a function of $k/k_c$. The classical Fisher information is calculated for the position (or momentum) measurement which is the optimal measurement saturating the Cram\'er-Rao bound (classical Fisher information is equal to the quantum Fisher information). Calculations made for $N=2$ atoms until $k/k_c = 0.9997$. In the bottom row, we show position correlations and how they change as a function of the unknown parameter (with fixed other parameters) which have to be measured for the optimal estimation.}
    \label{fig:fig2}
\end{figure}

So far we have calculated the quantum Fisher information which provides the ultimate bound on the precision through the Cram\'er-Rao bound $\Delta^2 \Omega \geq 1/\mathcal{I}_\Omega$. This bound can be saturated provided that an optimal measurement has been performed to extract the information from the system. In some cases, however, the optimal measurement might be not feasible in experiments. Therefore, we have to investigate what kind of measurement is optimal for the presented scheme. Fortunately, as the squeezed Fock states have real coefficients (up to a global phase), it can be argued that the optimal measurement amounts to measuring the position (or momentum) distribution of the atoms~\cite{2016wasakoptimal}. Since the expansion coefficients are real, no information is lost after transforming them into probabilities. In other words, no information about the unknown parameter is stored in the phase. This is briefly illustrated in figure~\ref{fig:fig2}, where we compare the quantum Fisher information with the (numerically calculated) classical Fisher information 
\begin{align}
    \mathcal{F}_\Omega = \sum_\eta \frac{1}{p(\eta|\Omega)}\left(\frac{\partial p(\eta|\Omega)}{\partial \Omega} \right)^2
\end{align}
where $p(\eta|\Omega)$ is the probability of measuring an outcome labeled by $\eta$ for a given value of an unknown parameter $\Omega$. 
Once the probability is known, the maximum likelihood estimator can be used to estimate the value of the unknown parameter~\cite{rossi2018mathematical}. Note that the entire probability is necessary for the optimal estimation. This might constitute a challenge for a macroscopic system which could be potentially overcome using time-reversal-based techniques~\cite{2016nosingleparticle,2016timereversalqm,vuletic2022timereversalmetrology} and interaction-based readouts~\cite{2016durlargeQFI,2017interactionbasedreadout}.

% For the excited state of a non-interacting Bose-Einstein condensate, the optimal measurement depends on time as the excited state performs an additional rotation in the phase space of the system. This might constitute another difficulty that could be overcome using an adaptive measurement scheme~\cite{wiseman_milburn_2009,2011Adaptivemetrology,2017Adaptivemetrologymarkovian,PRXQuantum.2.040301}.

%%%%%%%%%%%%%%%%%%%%%%%%%%%%%%%%%%%%%%%%%%%%%%%%%%%%%%%%%%%%%%%%%%%%%%%%%%%%
%%%%%%%%%%%%%%%%%%%%%%%%%%%%%%% CONCLUSIONS  %%%%%%%%%%%%%%%%%%%%%%%%%%%%%%%
%%%%%%%%%%%%%%%%%%%%%%%%%%%%%%%%%%%%%%%%%%%%%%%%%%%%%%%%%%%%%%%%%%%%%%%%%%%%

\section{Conclusions \& Outlook}
By making use of the observation that the measurement precision of a parameter in the Hamiltonian can grow with the size of the (parameter-dependent) Hilbert space, we propose to couple the motional degree of freedom of the atoms to their internal states. The effective spin-orbit coupling transfers the information about the unknown system parameter to be measured to atomic motion increasing thus the metrological capacity of the system. Note that the physics here is related to atom interferometry with momentum-dependent beam splitting. This leads to parameter-dependent energy level occupations of the trap oscillator states, which can be directly measured via the atomic density distribution including correlations. As a generic and largely solvable example, we consider a one-dimensional spin-polarized quantum gas of fermions and strongly interacting bosons that are coupled to the harmonic potential in which they reside. In this case, we can calculate the Fisher information explicitly and also identify a concrete measurement that can saturate the Cram\'er-Rao bound. Here it is simply the measurement of the position or momentum distribution, which is an elementary tool in the ultra-cold-gas-experiments toolbox. 

Our proposal is inspired by critical metrology~\cite{2020criticalwitkowska,Gietka2021adiabaticcritical,2021liuexperiment,2021criticalwitkowska,2022PRXQuantumcontinus,2022_Garbe_heisenbegkible,2022entropycritical,2022dingnature,Aybar2022criticalquantum,2022garbeexponetialsqueezing} and relies on adiabatic preparation of a stationary atomic state close to the quantum phase transition point towards stripe formation. This generates strongly squeezed center-of-mass excitations, which enables one to improve the precision of atomic frequency measurements and leads to a \emph{squeezed} Heisenberg scaling of precision. 

In an alternative setting, we identified a protocol that uses an entangled superposition of excited states of non-interacting spin-orbit coupled bosons. Here one can overcome the Heisenberg scaling of the precision by spreading the state over a much larger part of the Hilbert space compared to the ground state Bose-Einstein condensate. In practice, the precise preparation of such a state might be experimentally more challenging as there is no clear prescription on how to prepare a \emph{bosonized} version of a fermionic wavefunction. A possible solution might be using the BCS-BEC crossover~\cite{feschbach2010rmp} which was very recently employed to build a quantum engine~\cite{Menon2022bcsbec}.

To the best of our knowledge, we present the first ensemble quantum measurement protocols for fermions (not exploiting spin-squeezing~\cite{2022twoaxisemilia}) as well as a Tonks-Girardeau gas to reach the Heisenberg particle number scaling of precision. It is also the first proposal suggesting a principle possibility and a measurement procedure to overcome the Heisenberg scaling without resorting to $k$-body interactions~\cite{2011INTbeyondHL} to encode the parameter to measure into the initial state. We expect the proposal to be robust as it relies on adiabatic preparation of the ground state. As a central challenge in an experiment, we expect a finite ensemble temperature to be a major source of noise and error. For a finite temperature $T\neq 0$ incoherent excitations will lead to an increased number of center-of-mass excitations, which theoretically should lead to an increase of the quantum Fisher information~(see Appendix \ref{appendix:D} for details). %In this case the quantum Fisher information can be calculated to be 
% \begin{align}
%     \mathcal{I}_\Omega =  \frac{k^4}{k_c^4}\frac{1}{16\Omega^2 \left(1-\frac{k^2}{k_c^2}\right)^2} \frac{\tanh (\beta  \omega )+1}{ \tanh ^2\left(\frac{\beta  \omega }{2}\right)}.
% \end{align}

From a wider perspective, we feel that our proposal opens a new avenue in quantum metrology, where one employs extra degrees of freedom and in particular the center-of-mass excitations of the system---typically suppressed or neglected in calculations~\cite{pezze2018QM_RMP}---to allow for increased measurement precision. With the current extraordinary experimental control in complex quantum systems as in small quantum computing devices (NISQ) with ions or tweezer-trapped atoms, it should be already possible to test instances of this proposal at the level of few-body quantum systems. Here a transfer of quantum information between internal and external degrees of freedom is already routinely implemented to perform non-local gates~\cite{Soderberg_2010_review}. 

A promising possibility is to exploit the quench dynamics between two phases~\cite{Gietka2022understanding} or use the dynamical instability~\cite{2022mechanicalsqueezing} caused by an inverted harmonic oscillator dynamics~\cite{2021Inverted_Oscillator_Gietka,weiss2021large}. This should vastly decrease the time of the protocol and increase further the number of center-of-mass excitations required for improved precision. An interesting extension of this work might be spin-orbit coupled ultra-cold quantum gases trapped in cavity fields~\cite{2013helmutrmp,2019gietkagrav,2021Gietka_njp,2021farokhqed,2021amrcavitymetrology}. The cavity field constitutes another (macroscopic) harmonic oscillator to which the information about the unknown parameter could be coupled and further increase the precision. Also using entanglement between the two atomic levels in the form of spin squeezing and combining it with the squeezing of the center-of-mass motion might be a promising approach in quantum metrology. We defer detailed studies of all these possibilities to future investigation. 

%%%%%%%%%%%%%%%%%%%%%%%%%%%%%%%%%%%%%%%%%%%%%%%%%%%%%%%%%%%%%%%%%%%%%%%%%%%%
%%%%%%%%%%%%%%%%%%%%%%%%%%%%%  ACKNOWLEDGEMENTS  %%%%%%%%%%%%%%%%%%%%%%%%%%%
%%%%%%%%%%%%%%%%%%%%%%%%%%%%%%%%%%%%%%%%%%%%%%%%%%%%%%%%%%%%%%%%%%%%%%%%%%%%

\section{Acknowledgements}
K.G. is pleased to acknowledge Thom\'as Fogarty, Thomas Busch, Elo Cuestas, and Christoph Hotter for fruitful discussions. Simulations were performed using the open-source \textsc{QuantumOptics.jl} framework in \textsc{Julia}~\cite{kramer2018quantumoptics}. This work was supported by the Okinawa Institute of Science and Technology Graduate University and by the Lise-Meitner Fellowship M3304-N of the Austrian Science Fund (FWF).

%%%%%%%%%%%%%%%%%%%%%%%%%%%%%%%%%%%%%%%%%%%%%%%%%%%%%%%%%%%%%%%%%%%%%%%%%%%%
%%%%%%%%%%%%%%%%%%%%%%%%%%%%%%%  BIBLIOGRAPHY  %%%%%%%%%%%%%%%%%%%%%%%%%%%%%
%%%%%%%%%%%%%%%%%%%%%%%%%%%%%%%%%%%%%%%%%%%%%%%%%%%%%%%%%%%%%%%%%%%%%%%%%%%%

%apsrev4-2.bst 2019-01-14 (MD) hand-edited version of apsrev4-1.bst
%Control: key (0)
%Control: author (8) initials jnrlst
%Control: editor formatted (1) identically to author
%Control: production of article title (0) allowed
%Control: page (0) single
%Control: year (1) truncated
%Control: production of eprint (0) enabled
%

\onecolumngrid
 \appendix
 
%%%%%%%%%%%%%%%%%%%%%%%%%%%%%%%%%%%%%%%%%%%%%%%%%%%%%%%%%%%%%%%%%%%%%%%%%%%%
%%%%%%%%%%%%%%%%%%%%%%%%%%%%%%%% APPENDIX A %%%%%%%%%%%%%%%%%%%%%%%%%%%%%%%%
%%%%%%%%%%%%%%%%%%%%%%%%%%%%%%%%%%%%%%%%%%%%%%%%%%%%%%%%%%%%%%%%%%%%%%%%%%%%

\section{Derivation of the effective Hamiltonian}\label{appendix:A}
Here we derive the effective Hamiltonian from equation~\eqref{eq:effectiveH}. Starting from the single-particle Hamiltonian for a spin-orbit coupled two-level atom
\begin{align}
    \hat H = \frac{ \left(\hat p + k \hat \sigma_z\right)^2}{2m} + \frac{m \omega^2 \hat x^2}{2} + \frac{ \Omega}{2}  \hat \sigma_x,
\end{align}
we introduce first the annihilation and creation operators of the center-of-mass excitations
\begin{align}
    \hat a = \sqrt{\frac{m \omega}{2}}\left(\hat x + \frac{i}{m \omega} \hat p \right) \qquad \mathrm{and} \qquad  \hat a^\dagger = \sqrt{\frac{m \omega}{2}}\left(\hat x - \frac{i}{m \omega} \hat p \right).
\end{align}
upon inserting them into the Hamiltonian~\eqref{eq:socH} we obtain
\begin{align}
    \hat H = \omega\left(\hat a^\dagger\hat a +\frac{1}{2}\right) + i  \sqrt{\frac{ \omega}{2m}}\left( \hat a^\dagger - \hat a\right)  k \hat \sigma_z  + \frac{ \Omega}{2}  \hat \sigma_x,
\end{align}
By rotating the spin around the $y$ axis and rotating the phase space of the harmonic oscillator we finally get the Hamiltonian in the form of the quantum Rabi model (we dropped the constant $\omega/2$ factor)
\begin{align}\label{eq:intH}
    \hat H = \omega\hat a^\dagger\hat a  +   \sqrt{\frac{ \omega}{2m}}k  \left( \hat a^\dagger + \hat a\right) \hat \sigma_x  + \frac{ \Omega}{2}  \hat \sigma_z.
\end{align}
The rotation about the $y$ axis is not necessary and is introduced as a matter of convenience.

The Schrieffer-Wolff transformation
 \begin{equation}
      \hat U =\exp\left(i \frac{k}{k_c}\frac{\sqrt{\omega}}{2\sqrt{\Omega}}(\hat a +\hat a^\dagger){\hat \sigma_y} \right), 
 \end{equation}
 is a simultaneous rotation of the spin by an angle $\phi = \frac{k}{k_c}\frac{\sqrt{\omega}}{\sqrt{\Omega}} \left(\hat a^\dagger + \hat a\right)$ and displacement of the harmonic oscillator by $\alpha = i\frac{k}{k_c}\frac{\sqrt{\omega}}{2\sqrt{\Omega}} \hat \sigma_y$. Applying this transformation to the Hamiltonian~\eqref{eq:intH} yields
 \begin{align}
     \hat H =   \omega  \left(\hat a^\dagger-\alpha\right) \left(\hat a+\alpha\right) + \frac{\Omega}{2}  \left(\cos \phi \hat\sigma_z -\sin \phi \hat \sigma_x\right) + k\sqrt{\frac{ \omega}{2m}} \left(\hat a  + \hat a^\dagger\right) \left(\cos \phi \hat \sigma_x + \sin \phi \hat \sigma_z\right).
 \end{align}
 Assuming $\omega/\Omega \rightarrow 0$, we obtain Hamiltonian~\eqref{eq:effectiveH}
 \begin{align}
  \hat H \approx \omega \hat a^\dagger \hat a + \frac{\Omega}{2} \hat \sigma_z + k^2{\frac{ \omega}{2m \Omega}} (\hat a +  \hat a^\dagger)^2 \hat \sigma_z.
\end{align}
In this regime, the spin down is decoupled from spin up and the single-particle eigenstates of the system become
    \begin{align}
    |\psi_n(\xi) \rangle = \hat S(\xi) |n\rangle \otimes|\!\downarrow\,\rangle
\end{align}
with $\hat S(\xi) \equiv \exp\{(\xi/2)(\hat a^\dagger)^2-(\xi^*/2)\hat a^2\}$ being the squeeze operator where $\xi = -\frac{1}{4} \ln\{1-(k/k_c)^2\}$ and $k_c = \sqrt{\Omega \omega}$.

%%%%%%%%%%%%%%%%%%%%%%%%%%%%%%%%%%%%%%%%%%%%%%%%%%%%%%%%%%%%%%%%%%%%%%%%%%%%
%%%%%%%%%%%%%%%%%%%%%%%%%%%%%%%% APPENDIX B %%%%%%%%%%%%%%%%%%%%%%%%%%%%%%%%
%%%%%%%%%%%%%%%%%%%%%%%%%%%%%%%%%%%%%%%%%%%%%%%%%%%%%%%%%%%%%%%%%%%%%%%%%%%%

\section{Quantum Fisher information for the polarized gas of fermions}\label{appendix:B}
The fermionic ground-state wavefunction can be constructed out of single-particle eigenstates according to the Slater determinant
\begin{align}
  \Psi_\mathrm{F}(x_1,x_2,\ldots,x_N) \propto \mathrm{det}[\phi_i(x_j)], \qquad i,j =1,2,\ldots,N,
\end{align}
where $\phi_i(x_j)$ is the $i$th eigenstate of the single-particle Hamiltonian of the $j$th particle. Such a wavefunction contains $N!$ elements. Acting with the derivative operator on each term will give $N$ terms proportional to 
\begin{align}
    \hat h_j =  
    \frac{k^2}{k_c^2}\frac{i}{\frac{k^2}{k_c^2}-1}\frac{1}{8\Omega}\left(\hat a_j^\dagger \hat a_j^\dagger- \hat a_j\hat a_j  \right).
\end{align}
Then, the quantum Fisher information can be calculated as a variance of the local generator operator
\begin{align}
   \mathcal{I}_\Omega =  \Delta^2\left[\sum_{j=1}^N \hat h_j \right].
\end{align}
The first moment of $\sum_{j=1}^N \hat h_j$ is trivially 0 as only balanced terms where the number of creation and annihilation operators is equal will give non-zero contributions. Note that $\hat a^\dagger$ ($\hat a$) does not create (annihilate) a fermion, but a center-of-mass excitation. The calculation of the second moment is divided into two steps. First, we calculate the terms proportional to $\hat h_j^2$. Straightforward calculations yield
\begin{align}
\begin{split}
    \sum_{j=1}^N \langle \hat h_j^2\rangle = & \sum_{j=1}^{N} {\langle\Psi_\mathrm{F} |\left(\hat a_j^\dagger \hat a_j^\dagger \hat a_j \hat a_j + \hat a_j \hat a_j\hat a_j^\dagger \hat a_j^\dagger \right)|\Psi_\mathrm{F}\rangle}\frac{1}{64\Omega^2}\frac{k^4}{k_c^4}\frac{1}{\left(1-\frac{k^2}{k_c^2}\right)^2}  \\ = & \sum_{n=0}^{N-1} {(1+n+n^2)}\frac{1}{64\Omega^2}\frac{k^4}{k_c^4}\frac{1}{\left(1-\frac{k^2}{k_c^2}\right)^2} =  \frac{\frac{k^4}{k_c^4} N \left(N^2+2\right)}{24 \Omega ^2 \left(1-\frac{k^2}{k_c^2} \right)^2}.
    \end{split}
\end{align}
For the terms containing tensor products of $\hat h_j$ and $\hat h_k$, we get
\begin{align}\label{eq:minussum}
\begin{split}
   2 \sum_{j> k}^N \langle \hat h_j \hat h_k\rangle = &  2 \sum_{j> k}^N  {\langle\Psi_\mathrm{F} |\left(\hat a_j^\dagger \hat a_j^\dagger \hat a_k \hat a_k + \hat a_k \hat a_k\hat a_j^\dagger \hat a_j^\dagger \right)|\Psi_\mathrm{F}\rangle} \frac{1}{64\Omega^2}\frac{k^4}{k_c^4}\frac{1}{\left(1-\frac{k^2}{k_c^2}\right)^2}  \\ =  &  -\sum_{n=0}^{N-1} {n(n-1)}\frac{1}{8\Omega^2}\frac{k^4}{k_c^4}\frac{1}{\left(1-\frac{k^2}{k_c^2}\right)^2} = -\frac{\frac{k^4}{k_c^4} (N-2) (N-1) N}{24 \Omega ^2 \left(1 -\frac{k^2}{k_c^2}\right)^2}.
   \end{split}
\end{align}
By adding these terms we obtain equation~\eqref{eq:qfifermion} from the main text which exhibits Heisenberg scaling with respect to the number of atoms $N$
\begin{align} \label{eq:qfiN}
    \mathcal{I}_\Omega =  \frac{\frac{k^4}{k_c^4} N^2}{8 \Omega ^2 \left(1-\frac{k^2}{k_c^2} \right)^2}.
\end{align}
This scaling can be elevated to super Heisenberg scaling by noticing that the minus sign in front of the sum in equation ~\eqref{eq:minussum} derives from the anti-symmetric character of the wave function. For an excited state (but still an eigenstate) of a Bose-Einstein condensate in which atoms occupy coherently first $N$ modes similarly to fermions, 
% the system evolves according to
% \begin{align}
%     |\Psi(T)\rangle = \sum_{n=0}^{N-1}\frac{1}{\sqrt{N}} \exp\left[ i B_n \right]|n(T)\rangle,
% \end{align}
% where $B_n =\int_0^{k}\langle n|\partial_{k^\prime} |n\rangle \mathrm{d}k^\prime$ is the Berry phase of the $n$th eigenstate which is 0 for the same reason as the first moment of the local generator is 0.
the quantum Fisher information can be calculated to be
\begin{align}
    \mathcal{I}_\Omega = \frac{\frac{k^4}{k_c^4} \left(\frac{1}{2}N^3-\frac{6}{8}N^2+N\right)}{6 \Omega ^2 \left(1-\frac{k^2}{k_c^2} \right)^2}.
\end{align}
Note that such a state is extremely nonclassical as it represents $N$ squeezed harmonic oscillators entangled with each other.
% Plugging in the adiabatic ramp condition $k(t) = 2 k_c \frac{\sqrt{\gamma \omega t (\gamma \omega t +1)}}{1+2\gamma \omega t}$~\cite{2022criticalspeedup}, the quantum Fisher information becomes
% \begin{align}
%     \mathcal{I}_\Omega = \frac{\frac{k^4}{k_c^4} \left(\frac{1}{2}N^3-\frac{6}{8}N^2+N\right)}{6 \Omega ^2 \left(1-\frac{k^2}{k_c^2} \right)^2}.
% \end{align}

% Since this state is a product state, the extra $N$ improvement over the Heisenberg scaling is statistical in its nature. Such scaling is the consequence of the fact that even though particles are in a product state, they know that there is $N$ of them. In this sense, the scaling might seem artificial. However, as absolute precision---not its scaling---is the aim of metrology, using this state should lead to overcoming the single particle (standard quantum) limit and potentially the Heisenberg limit.

% Note that in the above calculations $\hat a$ and $\hat a^\dagger$ do not lead to annihilation or creation of a fermion, but of excitation of a (squeezed) harmonic oscillator. 
%%%%%%%%%%%%%%%%%%%%%%%%%%%%%%%%%%%%%%%%%%%%%%%%%%%%%%%%%%%%%%%%%%%%%%%%%%%%
%%%%%%%%%%%%%%%%%%%%%%%%%%%%%%%% APPENDIX C %%%%%%%%%%%%%%%%%%%%%%%%%%%%%%%%
%%%%%%%%%%%%%%%%%%%%%%%%%%%%%%%%%%%%%%%%%%%%%%%%%%%%%%%%%%%%%%%%%%%%%%%%%%%%

\section{Quantum Fisher information for the Tonks-Girardeau wavefunction}\label{appendix:C}
Here we calculate the quantum Fisher information for the Tonks-Girardeau gas. Using the spectral decomposition of the unity operator in the position basis we can rewrite the quantum Fisher information as
\begin{align}
    \mathcal{I}_\Omega = \langle \partial_\Omega \psi |\partial_\Omega \psi \rangle - |\langle \partial_\Omega \psi|\psi\rangle|^2 = \int_\mathbb{R} \left(\partial_\Omega \psi(x)\right)\left(\partial_\Omega \psi^*(x)\right) \mathrm{d}x- \int_\mathbb{R} \psi(x)\partial_\Omega \psi^*(x)\mathrm{d}x \int_\mathbb{R} \psi^*(y)\partial_\Omega \psi(y)\mathrm{d}y.
\end{align}

Since for squeezed Fock states of the harmonic oscillator $\psi_n(x) \equiv \langle x |\hat S(\xi)| n \rangle$ is real
and every term in the above expression contains even power of the wavefunction its sign does not play a role. Therefore the quantum Fisher information for $\psi(x)$ is the same as for $|\psi(x)|$. 

%%%%%%%%%%%%%%%%%%%%%%%%%%%%%%%%%%%%%%%%%%%%%%%%%%%%%%%%%%%%%%%%%%%%%%%%%%%%
%%%%%%%%%%%%%%%%%%%%%%%%%%%%%%%% APPENDIX D %%%%%%%%%%%%%%%%%%%%%%%%%%%%%%%%
%%%%%%%%%%%%%%%%%%%%%%%%%%%%%%%%%%%%%%%%%%%%%%%%%%%%%%%%%%%%%%%%%%%%%%%%%%%%

\section{Finite temperature considerations}\label{appendix:D}
Here we present calculations confirming that finite temperature might lead to the increase of the quantum Fisher information. For the sake of simplicity, we consider only one atom. The state of the system can be described using a density matrix operator (statistical mixture)
\begin{align}
    \hat \rho = \sum_{n=0}^{\infty} p_n |n\rangle \langle n|,
\end{align}
where $n$ labels the excited states and $p_n$ is given by
\begin{align}\label{eq:probab}
    p_n = \frac{1}{Z}\exp(- \beta E_n),
\end{align}
where $Z = \sum_n \exp(- \beta E_n)$ is the partition function (statistical sum) with $E_n$ being the (initial) energy of the $n$th excited state, and $\beta = 1/k_b T$ where $k_b$ is the Boltzmann constant and $T$ is the temperature. In order to calculate the quantum Fisher information for a statistical mixture, we have to use the following formula
\begin{align}
    \mathcal{I}_\Omega = 2 \sum_{n,m}\frac{(p_n-p_m)^2}{p_n+p_m}|\langle n| \hat h | m \rangle |^2,
\end{align}
where
\begin{align}
    \hat h = \frac{k^2}{k_c^2}\frac{i}{\frac{k^2}{k_c^2}-1}\frac{1}{8\Omega}\left( \hat a^\dagger \hat a^\dagger - \hat a\hat a\right),
\end{align}
and $|n\rangle$ is the $n$th Fock state. The only non-zero components will originate from states that differ by two excitations. Therefore, the quantum Fisher information can be rewritten as
\begin{align}
    \mathcal{I}_\Omega =  \frac{k^4}{k_c^4}\frac{1}{\left(1-\frac{k^2}{k_c^2}\right)^2}\frac{1}{32\Omega^2} \sum_{n,m}\frac{(p_n-p_m)^2}{p_n+p_m}\left[  (m+1)(m+2) \delta_{n,m+2} +  {m}(m-1) \delta_{n,m-2} \right].
\end{align}
After evaluating the Kronecker deltas we obtain
\begin{align}
    \mathcal{I}_\Omega =  \frac{k^4}{k_c^4}\frac{1}{\left(1-\frac{k^2}{k_c^2}\right)^2}\frac{1}{32\Omega^2} \left[\sum_{n=0}\frac{(p_{n+2}-p_n)^2}{p_{n+2}+p_n}  (n+1)(n+2)  +\sum_{n=2}\frac{(p_{n-2}-p_n)^2}{p_{n-2}+p_n}  {n}(n-1)  \right],
\end{align}
and finally, by redefining the second sum, we get
\begin{align}\label{eq:finalsum}
    \mathcal{I}_\Omega =  \frac{k^4}{k_c^4}\frac{1}{\left(1-\frac{k^2}{k_c^2}\right)^2}\frac{1}{16\Omega^2} \sum_{n=0}\frac{(p_{n+2}-p_n)^2}{p_{n+2}+p_n}  (n+1)(n+2).
\end{align}
Now, if only $p_0 \neq 0$, we obtain the result from reference~\cite{2020CriticalParis}
\begin{align}\label{eq:oldqfi}
    \mathcal{I}_\Omega =  \frac{k^4}{k_c^4}\frac{1}{\left(1-\frac{k^2}{k_c^2}\right)^2}\frac{1}{8\Omega^2}.
\end{align}
In any other case, the sum from equation~\eqref{eq:finalsum} will benefit from incoherently excited states leading to a larger quantum Fisher information than in equation~\eqref{eq:oldqfi}. In order to see it explicitly, we can rewrite the sum using the probabilities from equation~\eqref{eq:probab}
\begin{align}
    \sum_{n=0}\frac{2}{Z} \sinh (\beta  \omega) \tanh (\beta  \omega) e^{-\beta  \omega (n+1)}  (n+1)(n+2) = \frac{\tanh (\beta  \omega )+1}{ \tanh ^2\left(\frac{\beta  \omega }{2}\right)}.
\end{align}
For $\beta \gg \omega$ this expression is equal to 2 (hyperbolic tangents approach 1), and we recover the quantum Fisher information from equation~\eqref{eq:oldqfi}. Increasing the temperature (decreasing $\beta$) decreases the hyperbolic functions leading to enhancement of the quantum Fisher information. From a physical point of view, increasing the temperature increases the number of excitations which is a resource in quantum metrology. We expect that the same happens for a many-body quantum system.


\begin{thebibliography}{50}%
\makeatletter
\providecommand \@ifxundefined [1]{%
 \@ifx{#1\undefined}
}%
\providecommand \@ifnum [1]{%
 \ifnum #1\expandafter \@firstoftwo
 \else \expandafter \@secondoftwo
 \fi
}%
\providecommand \@ifx [1]{%
 \ifx #1\expandafter \@firstoftwo
 \else \expandafter \@secondoftwo
 \fi
}%
\providecommand \natexlab [1]{#1}%
\providecommand \enquote  [1]{``#1''}%
\providecommand \bibnamefont  [1]{#1}%
\providecommand \bibfnamefont [1]{#1}%
\providecommand \citenamefont [1]{#1}%
\providecommand \href@noop [0]{\@secondoftwo}%
\providecommand \href [0]{\begingroup \@sanitize@url \@href}%
\providecommand \@href[1]{\@@startlink{#1}\@@href}%
\providecommand \@@href[1]{\endgroup#1\@@endlink}%
\providecommand \@sanitize@url [0]{\catcode `\\12\catcode `\$12\catcode
  `\&12\catcode `\#12\catcode `\^12\catcode `\_12\catcode `\%12\relax}%
\providecommand \@@startlink[1]{}%
\providecommand \@@endlink[0]{}%
\providecommand \url  [0]{\begingroup\@sanitize@url \@url }%
\providecommand \@url [1]{\endgroup\@href {#1}{\urlprefix }}%
\providecommand \urlprefix  [0]{URL }%
\providecommand \Eprint [0]{\href }%
\providecommand \doibase [0]{https://doi.org/}%
\providecommand \selectlanguage [0]{\@gobble}%
\providecommand \bibinfo  [0]{\@secondoftwo}%
\providecommand \bibfield  [0]{\@secondoftwo}%
\providecommand \translation [1]{[#1]}%
\providecommand \BibitemOpen [0]{}%
\providecommand \bibitemStop [0]{}%
\providecommand \bibitemNoStop [0]{.\EOS\space}%
\providecommand \EOS [0]{\spacefactor3000\relax}%
\providecommand \BibitemShut  [1]{\csname bibitem#1\endcsname}%
\let\auto@bib@innerbib\@empty
%</preamble>
\bibitem [{\citenamefont {Giovannetti}\ \emph {et~al.}(2006)\citenamefont
  {Giovannetti}, \citenamefont {Lloyd},\ and\ \citenamefont
  {Maccone}}]{Giovanetti2006QuantumMetrology}%
  \BibitemOpen
  \bibfield  {author} {\bibinfo {author} {\bibfnamefont {V.}~\bibnamefont
  {Giovannetti}}, \bibinfo {author} {\bibfnamefont {S.}~\bibnamefont {Lloyd}},\
  and\ \bibinfo {author} {\bibfnamefont {L.}~\bibnamefont {Maccone}},\
  }\bibfield  {title} {\bibinfo {title} {Quantum metrology},\ }\href
  {https://doi.org/10.1103/PhysRevLett.96.010401} {\bibfield  {journal}
  {\bibinfo  {journal} {Phys. Rev. Lett.}\ }\textbf {\bibinfo {volume} {96}},\
  \bibinfo {pages} {010401} (\bibinfo {year} {2006})}\BibitemShut {NoStop}%
\bibitem [{\citenamefont {Giovannetti}\ \emph {et~al.}(2011)\citenamefont
  {Giovannetti}, \citenamefont {Lloyd},\ and\ \citenamefont
  {Maccone}}]{Giovanetti2011advancesinQM}%
  \BibitemOpen
  \bibfield  {author} {\bibinfo {author} {\bibfnamefont {V.}~\bibnamefont
  {Giovannetti}}, \bibinfo {author} {\bibfnamefont {S.}~\bibnamefont {Lloyd}},\
  and\ \bibinfo {author} {\bibfnamefont {L.}~\bibnamefont {Maccone}},\
  }\bibfield  {title} {\bibinfo {title} {Advances in quantum metrology},\
  }\href {https://doi.org/10.1038/nphoton.2011.35} {\bibfield  {journal}
  {\bibinfo  {journal} {Nature Photonics}\ }\textbf {\bibinfo {volume} {5}},\
  \bibinfo {pages} {222} (\bibinfo {year} {2011})}\BibitemShut {NoStop}%
\bibitem [{\citenamefont {Pezz\`e}\ \emph {et~al.}(2018)\citenamefont
  {Pezz\`e}, \citenamefont {Smerzi}, \citenamefont {Oberthaler}, \citenamefont
  {Schmied},\ and\ \citenamefont {Treutlein}}]{pezze2018QM_RMP}%
  \BibitemOpen
  \bibfield  {author} {\bibinfo {author} {\bibfnamefont {L.}~\bibnamefont
  {Pezz\`e}}, \bibinfo {author} {\bibfnamefont {A.}~\bibnamefont {Smerzi}},
  \bibinfo {author} {\bibfnamefont {M.~K.}\ \bibnamefont {Oberthaler}},
  \bibinfo {author} {\bibfnamefont {R.}~\bibnamefont {Schmied}},\ and\ \bibinfo
  {author} {\bibfnamefont {P.}~\bibnamefont {Treutlein}},\ }\bibfield  {title}
  {\bibinfo {title} {Quantum metrology with nonclassical states of atomic
  ensembles},\ }\href {https://doi.org/10.1103/RevModPhys.90.035005} {\bibfield
   {journal} {\bibinfo  {journal} {Rev. Mod. Phys.}\ }\textbf {\bibinfo
  {volume} {90}},\ \bibinfo {pages} {035005} (\bibinfo {year}
  {2018})}\BibitemShut {NoStop}%
\bibitem [{\citenamefont {Wineland}\ \emph {et~al.}(1992)\citenamefont
  {Wineland}, \citenamefont {Bollinger}, \citenamefont {Itano}, \citenamefont
  {Moore},\ and\ \citenamefont {Heinzen}}]{wineland1992squeezing}%
  \BibitemOpen
  \bibfield  {author} {\bibinfo {author} {\bibfnamefont {D.~J.}\ \bibnamefont
  {Wineland}}, \bibinfo {author} {\bibfnamefont {J.~J.}\ \bibnamefont
  {Bollinger}}, \bibinfo {author} {\bibfnamefont {W.~M.}\ \bibnamefont
  {Itano}}, \bibinfo {author} {\bibfnamefont {F.~L.}\ \bibnamefont {Moore}},\
  and\ \bibinfo {author} {\bibfnamefont {D.~J.}\ \bibnamefont {Heinzen}},\
  }\bibfield  {title} {\bibinfo {title} {Spin squeezing and reduced quantum
  noise in spectroscopy},\ }\href {https://doi.org/10.1103/PhysRevA.46.R6797}
  {\bibfield  {journal} {\bibinfo  {journal} {Phys. Rev. A}\ }\textbf {\bibinfo
  {volume} {46}},\ \bibinfo {pages} {R6797} (\bibinfo {year}
  {1992})}\BibitemShut {NoStop}%
\bibitem [{\citenamefont {Ma}\ \emph {et~al.}(2011)\citenamefont {Ma},
  \citenamefont {Wang}, \citenamefont {Sun},\ and\ \citenamefont
  {Nori}}]{MA2011quantumSS}%
  \BibitemOpen
  \bibfield  {author} {\bibinfo {author} {\bibfnamefont {J.}~\bibnamefont
  {Ma}}, \bibinfo {author} {\bibfnamefont {X.}~\bibnamefont {Wang}}, \bibinfo
  {author} {\bibfnamefont {C.}~\bibnamefont {Sun}},\ and\ \bibinfo {author}
  {\bibfnamefont {F.}~\bibnamefont {Nori}},\ }\bibfield  {title} {\bibinfo
  {title} {Quantum spin squeezing},\ }\href
  {https://doi.org/https://doi.org/10.1016/j.physrep.2011.08.003} {\bibfield
  {journal} {\bibinfo  {journal} {Physics Reports}\ }\textbf {\bibinfo {volume}
  {509}},\ \bibinfo {pages} {89} (\bibinfo {year} {2011})}\BibitemShut
  {NoStop}%
\bibitem [{\citenamefont {Demkowicz-Dobrza{\'n}ski}\ \emph
  {et~al.}(2012)\citenamefont {Demkowicz-Dobrza{\'n}ski}, \citenamefont
  {Ko{\l}ody{\'n}ski},\ and\ \citenamefont {Gu{\c t}{\u a}}}]{2012rafaljan}%
  \BibitemOpen
  \bibfield  {author} {\bibinfo {author} {\bibfnamefont {R.}~\bibnamefont
  {Demkowicz-Dobrza{\'n}ski}}, \bibinfo {author} {\bibfnamefont
  {J.}~\bibnamefont {Ko{\l}ody{\'n}ski}},\ and\ \bibinfo {author}
  {\bibfnamefont {M.}~\bibnamefont {Gu{\c t}{\u a}}},\ }\bibfield  {title}
  {\bibinfo {title} {The elusive heisenberg limit in quantum-enhanced
  metrology},\ }\href {https://doi.org/10.1038/ncomms2067} {\bibfield
  {journal} {\bibinfo  {journal} {Nature Communications}\ }\textbf {\bibinfo
  {volume} {3}},\ \bibinfo {pages} {1063} (\bibinfo {year} {2012})}\BibitemShut
  {NoStop}%
\bibitem [{\citenamefont {Zhang}\ \emph {et~al.}(2016)\citenamefont {Zhang},
  \citenamefont {Mossman}, \citenamefont {Busch}, \citenamefont {Engels},\ and\
  \citenamefont {Zhang}}]{2016propertiessocbec}%
  \BibitemOpen
  \bibfield  {author} {\bibinfo {author} {\bibfnamefont {Y.}~\bibnamefont
  {Zhang}}, \bibinfo {author} {\bibfnamefont {M.~E.}\ \bibnamefont {Mossman}},
  \bibinfo {author} {\bibfnamefont {T.}~\bibnamefont {Busch}}, \bibinfo
  {author} {\bibfnamefont {P.}~\bibnamefont {Engels}},\ and\ \bibinfo {author}
  {\bibfnamefont {C.}~\bibnamefont {Zhang}},\ }\bibfield  {title} {\bibinfo
  {title} {Properties of spin--orbit-coupled bose--einstein condensates},\
  }\href {https://doi.org/10.1007/s11467-016-0560-y} {\bibfield  {journal}
  {\bibinfo  {journal} {Front. Phys.}\ }\textbf {\bibinfo {volume} {11}},\
  \bibinfo {pages} {118103} (\bibinfo {year} {2016})}\BibitemShut {NoStop}%
\bibitem [{\citenamefont {Lin}\ \emph {et~al.}(2011)\citenamefont {Lin},
  \citenamefont {Jim{\'e}nez-Garc{\'\i}a},\ and\ \citenamefont
  {Spielman}}]{2011socbec}%
  \BibitemOpen
  \bibfield  {author} {\bibinfo {author} {\bibfnamefont {Y.~J.}\ \bibnamefont
  {Lin}}, \bibinfo {author} {\bibfnamefont {K.}~\bibnamefont
  {Jim{\'e}nez-Garc{\'\i}a}},\ and\ \bibinfo {author} {\bibfnamefont {I.~B.}\
  \bibnamefont {Spielman}},\ }\bibfield  {title} {\bibinfo {title}
  {Spin--orbit-coupled bose--einstein condensates},\ }\href
  {https://doi.org/10.1038/nature09887} {\bibfield  {journal} {\bibinfo
  {journal} {Nature}\ }\textbf {\bibinfo {volume} {471}},\ \bibinfo {pages}
  {83} (\bibinfo {year} {2011})}\BibitemShut {NoStop}%
\bibitem [{\citenamefont {Li}\ \emph {et~al.}(2016)\citenamefont {Li},
  \citenamefont {Huang}, \citenamefont {Shteynas}, \citenamefont {Burchesky},
  \citenamefont {Top}, \citenamefont {Su}, \citenamefont {Lee}, \citenamefont
  {Jamison},\ and\ \citenamefont {Ketterle}}]{2016socbecketterle}%
  \BibitemOpen
  \bibfield  {author} {\bibinfo {author} {\bibfnamefont {J.}~\bibnamefont
  {Li}}, \bibinfo {author} {\bibfnamefont {W.}~\bibnamefont {Huang}}, \bibinfo
  {author} {\bibfnamefont {B.}~\bibnamefont {Shteynas}}, \bibinfo {author}
  {\bibfnamefont {S.}~\bibnamefont {Burchesky}}, \bibinfo {author}
  {\bibfnamefont {F.~i. m. c. b. u. i. e. i.~f.}\ \bibnamefont {Top}}, \bibinfo
  {author} {\bibfnamefont {E.}~\bibnamefont {Su}}, \bibinfo {author}
  {\bibfnamefont {J.}~\bibnamefont {Lee}}, \bibinfo {author} {\bibfnamefont
  {A.~O.}\ \bibnamefont {Jamison}},\ and\ \bibinfo {author} {\bibfnamefont
  {W.}~\bibnamefont {Ketterle}},\ }\bibfield  {title} {\bibinfo {title}
  {Spin-orbit coupling and spin textures in optical superlattices},\ }\href
  {https://doi.org/10.1103/PhysRevLett.117.185301} {\bibfield  {journal}
  {\bibinfo  {journal} {Phys. Rev. Lett.}\ }\textbf {\bibinfo {volume} {117}},\
  \bibinfo {pages} {185301} (\bibinfo {year} {2016})}\BibitemShut {NoStop}%
\bibitem [{\citenamefont {Hwang}\ \emph {et~al.}(2015)\citenamefont {Hwang},
  \citenamefont {Puebla},\ and\ \citenamefont {Plenio}}]{2015qpt_qrm}%
  \BibitemOpen
  \bibfield  {author} {\bibinfo {author} {\bibfnamefont {M.-J.}\ \bibnamefont
  {Hwang}}, \bibinfo {author} {\bibfnamefont {R.}~\bibnamefont {Puebla}},\ and\
  \bibinfo {author} {\bibfnamefont {M.~B.}\ \bibnamefont {Plenio}},\ }\bibfield
   {title} {\bibinfo {title} {Quantum phase transition and universal dynamics
  in the rabi model},\ }\href {https://doi.org/10.1103/PhysRevLett.115.180404}
  {\bibfield  {journal} {\bibinfo  {journal} {Phys. Rev. Lett.}\ }\textbf
  {\bibinfo {volume} {115}},\ \bibinfo {pages} {180404} (\bibinfo {year}
  {2015})}\BibitemShut {NoStop}%
\bibitem [{\citenamefont {Garbe}\ \emph {et~al.}(2020)\citenamefont {Garbe},
  \citenamefont {Bina}, \citenamefont {Keller}, \citenamefont {Paris},\ and\
  \citenamefont {Felicetti}}]{2020CriticalParis}%
  \BibitemOpen
  \bibfield  {author} {\bibinfo {author} {\bibfnamefont {L.}~\bibnamefont
  {Garbe}}, \bibinfo {author} {\bibfnamefont {M.}~\bibnamefont {Bina}},
  \bibinfo {author} {\bibfnamefont {A.}~\bibnamefont {Keller}}, \bibinfo
  {author} {\bibfnamefont {M.~G.~A.}\ \bibnamefont {Paris}},\ and\ \bibinfo
  {author} {\bibfnamefont {S.}~\bibnamefont {Felicetti}},\ }\bibfield  {title}
  {\bibinfo {title} {Critical quantum metrology with a finite-component quantum
  phase transition},\ }\href {https://doi.org/10.1103/PhysRevLett.124.120504}
  {\bibfield  {journal} {\bibinfo  {journal} {Phys. Rev. Lett.}\ }\textbf
  {\bibinfo {volume} {124}},\ \bibinfo {pages} {120504} (\bibinfo {year}
  {2020})}\BibitemShut {NoStop}%
\bibitem [{\citenamefont {Braunstein}\ and\ \citenamefont
  {Caves}(1994)}]{1994distancestates}%
  \BibitemOpen
  \bibfield  {author} {\bibinfo {author} {\bibfnamefont {S.~L.}\ \bibnamefont
  {Braunstein}}\ and\ \bibinfo {author} {\bibfnamefont {C.~M.}\ \bibnamefont
  {Caves}},\ }\bibfield  {title} {\bibinfo {title} {Statistical distance and
  the geometry of quantum states},\ }\href
  {https://doi.org/10.1103/PhysRevLett.72.3439} {\bibfield  {journal} {\bibinfo
   {journal} {Phys. Rev. Lett.}\ }\textbf {\bibinfo {volume} {72}},\ \bibinfo
  {pages} {3439} (\bibinfo {year} {1994})}\BibitemShut {NoStop}%
\bibitem [{\citenamefont {Pezz\'e}\ and\ \citenamefont
  {Smerzi}(2009)}]{2009pezzesmerzientHL}%
  \BibitemOpen
  \bibfield  {author} {\bibinfo {author} {\bibfnamefont {L.}~\bibnamefont
  {Pezz\'e}}\ and\ \bibinfo {author} {\bibfnamefont {A.}~\bibnamefont
  {Smerzi}},\ }\bibfield  {title} {\bibinfo {title} {Entanglement, nonlinear
  dynamics, and the heisenberg limit},\ }\href
  {https://doi.org/10.1103/PhysRevLett.102.100401} {\bibfield  {journal}
  {\bibinfo  {journal} {Phys. Rev. Lett.}\ }\textbf {\bibinfo {volume} {102}},\
  \bibinfo {pages} {100401} (\bibinfo {year} {2009})}\BibitemShut {NoStop}%
\bibitem [{\citenamefont {Gietka}(2022{\natexlab{a}})}]{2022criticalspeedup}%
  \BibitemOpen
  \bibfield  {author} {\bibinfo {author} {\bibfnamefont {K.}~\bibnamefont
  {Gietka}},\ }\bibfield  {title} {\bibinfo {title} {Squeezing by critical
  speeding up: Applications in quantum metrology},\ }\href
  {https://doi.org/10.1103/PhysRevA.105.042620} {\bibfield  {journal} {\bibinfo
   {journal} {Phys. Rev. A}\ }\textbf {\bibinfo {volume} {105}},\ \bibinfo
  {pages} {042620} (\bibinfo {year} {2022}{\natexlab{a}})}\BibitemShut
  {NoStop}%
\bibitem [{\citenamefont {Tonks}(1936)}]{1936tonks}%
  \BibitemOpen
  \bibfield  {author} {\bibinfo {author} {\bibfnamefont {L.}~\bibnamefont
  {Tonks}},\ }\bibfield  {title} {\bibinfo {title} {The complete equation of
  state of one, two and three-dimensional gases of hard elastic spheres},\
  }\href {https://doi.org/10.1103/PhysRev.50.955} {\bibfield  {journal}
  {\bibinfo  {journal} {Phys. Rev.}\ }\textbf {\bibinfo {volume} {50}},\
  \bibinfo {pages} {955} (\bibinfo {year} {1936})}\BibitemShut {NoStop}%
\bibitem [{\citenamefont {Girardeau}(1960)}]{1960girardeau}%
  \BibitemOpen
  \bibfield  {author} {\bibinfo {author} {\bibfnamefont {M.}~\bibnamefont
  {Girardeau}},\ }\bibfield  {title} {\bibinfo {title} {Relationship between
  systems of impenetrable bosons and fermions in one dimension},\ }\href
  {https://doi.org/10.1063/1.1703687} {\bibfield  {journal} {\bibinfo
  {journal} {Journal of Mathematical Physics}\ }\textbf {\bibinfo {volume}
  {1}},\ \bibinfo {pages} {516} (\bibinfo {year} {1960})},\ \Eprint
  {https://arxiv.org/abs/https://doi.org/10.1063/1.1703687}
  {https://doi.org/10.1063/1.1703687} \BibitemShut {NoStop}%
\bibitem [{\citenamefont {Paredes}\ \emph {et~al.}(2004)\citenamefont
  {Paredes}, \citenamefont {Widera}, \citenamefont {Murg}, \citenamefont
  {Mandel}, \citenamefont {F{\"o}lling}, \citenamefont {Cirac}, \citenamefont
  {Shlyapnikov}, \citenamefont {H{\"a}nsch},\ and\ \citenamefont
  {Bloch}}]{2004TGgas}%
  \BibitemOpen
  \bibfield  {author} {\bibinfo {author} {\bibfnamefont {B.}~\bibnamefont
  {Paredes}}, \bibinfo {author} {\bibfnamefont {A.}~\bibnamefont {Widera}},
  \bibinfo {author} {\bibfnamefont {V.}~\bibnamefont {Murg}}, \bibinfo {author}
  {\bibfnamefont {O.}~\bibnamefont {Mandel}}, \bibinfo {author} {\bibfnamefont
  {S.}~\bibnamefont {F{\"o}lling}}, \bibinfo {author} {\bibfnamefont
  {I.}~\bibnamefont {Cirac}}, \bibinfo {author} {\bibfnamefont {G.~V.}\
  \bibnamefont {Shlyapnikov}}, \bibinfo {author} {\bibfnamefont {T.~W.}\
  \bibnamefont {H{\"a}nsch}},\ and\ \bibinfo {author} {\bibfnamefont
  {I.}~\bibnamefont {Bloch}},\ }\bibfield  {title} {\bibinfo {title}
  {Tonks--girardeau gas of ultracold atoms in an optical lattice},\ }\href
  {https://doi.org/10.1038/nature02530} {\bibfield  {journal} {\bibinfo
  {journal} {Nature}\ }\textbf {\bibinfo {volume} {429}},\ \bibinfo {pages}
  {277} (\bibinfo {year} {2004})}\BibitemShut {NoStop}%
\bibitem [{\citenamefont {Gietka}(2022{\natexlab{b}})}]{2022commetrology}%
  \BibitemOpen
  \bibfield  {author} {\bibinfo {author} {\bibfnamefont {K.}~\bibnamefont
  {Gietka}},\ }\bibfield  {title} {\bibinfo {title} {Harnessing center-of-mass
  excitations in quantum metrology},\ }\href
  {https://doi.org/10.1103/PhysRevResearch.4.043074} {\bibfield  {journal}
  {\bibinfo  {journal} {Phys. Rev. Research}\ }\textbf {\bibinfo {volume}
  {4}},\ \bibinfo {pages} {043074} (\bibinfo {year}
  {2022}{\natexlab{b}})}\BibitemShut {NoStop}%
\bibitem [{\citenamefont {Wasak}\ \emph {et~al.}(2016)\citenamefont {Wasak},
  \citenamefont {Smerzi}, \citenamefont {Pezz{\'e}},\ and\ \citenamefont
  {Chwede{\'n}czuk}}]{2016wasakoptimal}%
  \BibitemOpen
  \bibfield  {author} {\bibinfo {author} {\bibfnamefont {T.}~\bibnamefont
  {Wasak}}, \bibinfo {author} {\bibfnamefont {A.}~\bibnamefont {Smerzi}},
  \bibinfo {author} {\bibfnamefont {L.}~\bibnamefont {Pezz{\'e}}},\ and\
  \bibinfo {author} {\bibfnamefont {J.}~\bibnamefont {Chwede{\'n}czuk}},\
  }\bibfield  {title} {\bibinfo {title} {Optimal measurements in phase
  estimation: simple examples},\ }\href
  {https://doi.org/10.1007/s11128-016-1248-5} {\bibfield  {journal} {\bibinfo
  {journal} {Quantum Information Processing}\ }\textbf {\bibinfo {volume}
  {15}},\ \bibinfo {pages} {2231} (\bibinfo {year} {2016})}\BibitemShut
  {NoStop}%
\bibitem [{\citenamefont {Rossi}(2018)}]{rossi2018mathematical}%
  \BibitemOpen
  \bibfield  {author} {\bibinfo {author} {\bibfnamefont {R.~J.}\ \bibnamefont
  {Rossi}},\ }\href@noop {} {\emph {\bibinfo {title} {Mathematical statistics:
  an introduction to likelihood based inference}}}\ (\bibinfo  {publisher}
  {John Wiley \& Sons},\ \bibinfo {year} {2018})\BibitemShut {NoStop}%
\bibitem [{\citenamefont {Davis}\ \emph {et~al.}(2016)\citenamefont {Davis},
  \citenamefont {Bentsen},\ and\ \citenamefont
  {Schleier-Smith}}]{2016nosingleparticle}%
  \BibitemOpen
  \bibfield  {author} {\bibinfo {author} {\bibfnamefont {E.}~\bibnamefont
  {Davis}}, \bibinfo {author} {\bibfnamefont {G.}~\bibnamefont {Bentsen}},\
  and\ \bibinfo {author} {\bibfnamefont {M.}~\bibnamefont {Schleier-Smith}},\
  }\bibfield  {title} {\bibinfo {title} {Approaching the heisenberg limit
  without single-particle detection},\ }\href
  {https://doi.org/10.1103/PhysRevLett.116.053601} {\bibfield  {journal}
  {\bibinfo  {journal} {Phys. Rev. Lett.}\ }\textbf {\bibinfo {volume} {116}},\
  \bibinfo {pages} {053601} (\bibinfo {year} {2016})}\BibitemShut {NoStop}%
\bibitem [{\citenamefont {Linnemann}\ \emph {et~al.}(2016)\citenamefont
  {Linnemann}, \citenamefont {Strobel}, \citenamefont {Muessel}, \citenamefont
  {Schulz}, \citenamefont {Lewis-Swan}, \citenamefont {Kheruntsyan},\ and\
  \citenamefont {Oberthaler}}]{2016timereversalqm}%
  \BibitemOpen
  \bibfield  {author} {\bibinfo {author} {\bibfnamefont {D.}~\bibnamefont
  {Linnemann}}, \bibinfo {author} {\bibfnamefont {H.}~\bibnamefont {Strobel}},
  \bibinfo {author} {\bibfnamefont {W.}~\bibnamefont {Muessel}}, \bibinfo
  {author} {\bibfnamefont {J.}~\bibnamefont {Schulz}}, \bibinfo {author}
  {\bibfnamefont {R.~J.}\ \bibnamefont {Lewis-Swan}}, \bibinfo {author}
  {\bibfnamefont {K.~V.}\ \bibnamefont {Kheruntsyan}},\ and\ \bibinfo {author}
  {\bibfnamefont {M.~K.}\ \bibnamefont {Oberthaler}},\ }\bibfield  {title}
  {\bibinfo {title} {Quantum-enhanced sensing based on time reversal of
  nonlinear dynamics},\ }\href {https://doi.org/10.1103/PhysRevLett.117.013001}
  {\bibfield  {journal} {\bibinfo  {journal} {Phys. Rev. Lett.}\ }\textbf
  {\bibinfo {volume} {117}},\ \bibinfo {pages} {013001} (\bibinfo {year}
  {2016})}\BibitemShut {NoStop}%
\bibitem [{\citenamefont {Colombo}\ \emph {et~al.}(2022)\citenamefont
  {Colombo}, \citenamefont {Pedrozo-Pe{\~n}afiel}, \citenamefont {Adiyatullin},
  \citenamefont {Li}, \citenamefont {Mendez}, \citenamefont {Shu},\ and\
  \citenamefont {Vuleti{\'c}}}]{vuletic2022timereversalmetrology}%
  \BibitemOpen
  \bibfield  {author} {\bibinfo {author} {\bibfnamefont {S.}~\bibnamefont
  {Colombo}}, \bibinfo {author} {\bibfnamefont {E.}~\bibnamefont
  {Pedrozo-Pe{\~n}afiel}}, \bibinfo {author} {\bibfnamefont {A.~F.}\
  \bibnamefont {Adiyatullin}}, \bibinfo {author} {\bibfnamefont
  {Z.}~\bibnamefont {Li}}, \bibinfo {author} {\bibfnamefont {E.}~\bibnamefont
  {Mendez}}, \bibinfo {author} {\bibfnamefont {C.}~\bibnamefont {Shu}},\ and\
  \bibinfo {author} {\bibfnamefont {V.}~\bibnamefont {Vuleti{\'c}}},\
  }\bibfield  {title} {\bibinfo {title} {Time-reversal-based quantum metrology
  with many-body entangled states},\ }\href
  {https://doi.org/10.1038/s41567-022-01653-5} {\bibfield  {journal} {\bibinfo
  {journal} {Nature Physics}\ }\textbf {\bibinfo {volume} {18}},\ \bibinfo
  {pages} {925} (\bibinfo {year} {2022})}\BibitemShut {NoStop}%
\bibitem [{\citenamefont {Fr\"owis}\ \emph {et~al.}(2016)\citenamefont
  {Fr\"owis}, \citenamefont {Sekatski},\ and\ \citenamefont
  {D\"ur}}]{2016durlargeQFI}%
  \BibitemOpen
  \bibfield  {author} {\bibinfo {author} {\bibfnamefont {F.}~\bibnamefont
  {Fr\"owis}}, \bibinfo {author} {\bibfnamefont {P.}~\bibnamefont {Sekatski}},\
  and\ \bibinfo {author} {\bibfnamefont {W.}~\bibnamefont {D\"ur}},\ }\bibfield
   {title} {\bibinfo {title} {Detecting large quantum fisher information with
  finite measurement precision},\ }\href
  {https://doi.org/10.1103/PhysRevLett.116.090801} {\bibfield  {journal}
  {\bibinfo  {journal} {Phys. Rev. Lett.}\ }\textbf {\bibinfo {volume} {116}},\
  \bibinfo {pages} {090801} (\bibinfo {year} {2016})}\BibitemShut {NoStop}%
\bibitem [{\citenamefont {Nolan}\ \emph {et~al.}(2017)\citenamefont {Nolan},
  \citenamefont {Szigeti},\ and\ \citenamefont
  {Haine}}]{2017interactionbasedreadout}%
  \BibitemOpen
  \bibfield  {author} {\bibinfo {author} {\bibfnamefont {S.~P.}\ \bibnamefont
  {Nolan}}, \bibinfo {author} {\bibfnamefont {S.~S.}\ \bibnamefont {Szigeti}},\
  and\ \bibinfo {author} {\bibfnamefont {S.~A.}\ \bibnamefont {Haine}},\
  }\bibfield  {title} {\bibinfo {title} {Optimal and robust quantum metrology
  using interaction-based readouts},\ }\href
  {https://doi.org/10.1103/PhysRevLett.119.193601} {\bibfield  {journal}
  {\bibinfo  {journal} {Phys. Rev. Lett.}\ }\textbf {\bibinfo {volume} {119}},\
  \bibinfo {pages} {193601} (\bibinfo {year} {2017})}\BibitemShut {NoStop}%
\bibitem [{\citenamefont {Mirkhalaf}\ \emph {et~al.}(2020)\citenamefont
  {Mirkhalaf}, \citenamefont {Witkowska},\ and\ \citenamefont
  {Lepori}}]{2020criticalwitkowska}%
  \BibitemOpen
  \bibfield  {author} {\bibinfo {author} {\bibfnamefont {S.~S.}\ \bibnamefont
  {Mirkhalaf}}, \bibinfo {author} {\bibfnamefont {E.}~\bibnamefont
  {Witkowska}},\ and\ \bibinfo {author} {\bibfnamefont {L.}~\bibnamefont
  {Lepori}},\ }\bibfield  {title} {\bibinfo {title} {Supersensitive quantum
  sensor based on criticality in an antiferromagnetic spinor condensate},\
  }\href {https://doi.org/10.1103/PhysRevA.101.043609} {\bibfield  {journal}
  {\bibinfo  {journal} {Phys. Rev. A}\ }\textbf {\bibinfo {volume} {101}},\
  \bibinfo {pages} {043609} (\bibinfo {year} {2020})}\BibitemShut {NoStop}%
\bibitem [{\citenamefont {Gietka}\ \emph
  {et~al.}(2021{\natexlab{a}})\citenamefont {Gietka}, \citenamefont {Metz},
  \citenamefont {Keller},\ and\ \citenamefont
  {Li}}]{Gietka2021adiabaticcritical}%
  \BibitemOpen
  \bibfield  {author} {\bibinfo {author} {\bibfnamefont {K.}~\bibnamefont
  {Gietka}}, \bibinfo {author} {\bibfnamefont {F.}~\bibnamefont {Metz}},
  \bibinfo {author} {\bibfnamefont {T.}~\bibnamefont {Keller}},\ and\ \bibinfo
  {author} {\bibfnamefont {J.}~\bibnamefont {Li}},\ }\bibfield  {title}
  {\bibinfo {title} {Adiabatic critical quantum metrology cannot reach the
  {H}eisenberg limit even when shortcuts to adiabaticity are applied},\ }\href
  {https://doi.org/10.22331/q-2021-07-01-489} {\bibfield  {journal} {\bibinfo
  {journal} {{Quantum}}\ }\textbf {\bibinfo {volume} {5}},\ \bibinfo {pages}
  {489} (\bibinfo {year} {2021}{\natexlab{a}})}\BibitemShut {NoStop}%
\bibitem [{\citenamefont {Liu}\ \emph {et~al.}(2021)\citenamefont {Liu},
  \citenamefont {Chen}, \citenamefont {Jiang}, \citenamefont {Yang},
  \citenamefont {Wu}, \citenamefont {Li}, \citenamefont {Yuan}, \citenamefont
  {Peng},\ and\ \citenamefont {Du}}]{2021liuexperiment}%
  \BibitemOpen
  \bibfield  {author} {\bibinfo {author} {\bibfnamefont {R.}~\bibnamefont
  {Liu}}, \bibinfo {author} {\bibfnamefont {Y.}~\bibnamefont {Chen}}, \bibinfo
  {author} {\bibfnamefont {M.}~\bibnamefont {Jiang}}, \bibinfo {author}
  {\bibfnamefont {X.}~\bibnamefont {Yang}}, \bibinfo {author} {\bibfnamefont
  {Z.}~\bibnamefont {Wu}}, \bibinfo {author} {\bibfnamefont {Y.}~\bibnamefont
  {Li}}, \bibinfo {author} {\bibfnamefont {H.}~\bibnamefont {Yuan}}, \bibinfo
  {author} {\bibfnamefont {X.}~\bibnamefont {Peng}},\ and\ \bibinfo {author}
  {\bibfnamefont {J.}~\bibnamefont {Du}},\ }\bibfield  {title} {\bibinfo
  {title} {Experimental critical quantum metrology with the heisenberg
  scaling},\ }\href {https://doi.org/10.1038/s41534-021-00507-x} {\bibfield
  {journal} {\bibinfo  {journal} {npj Quantum Information}\ }\textbf {\bibinfo
  {volume} {7}},\ \bibinfo {pages} {170} (\bibinfo {year} {2021})}\BibitemShut
  {NoStop}%
\bibitem [{\citenamefont {Mirkhalaf}\ \emph {et~al.}(2021)\citenamefont
  {Mirkhalaf}, \citenamefont {Benedicto~Orenes}, \citenamefont {Mitchell},\
  and\ \citenamefont {Witkowska}}]{2021criticalwitkowska}%
  \BibitemOpen
  \bibfield  {author} {\bibinfo {author} {\bibfnamefont {S.~S.}\ \bibnamefont
  {Mirkhalaf}}, \bibinfo {author} {\bibfnamefont {D.}~\bibnamefont
  {Benedicto~Orenes}}, \bibinfo {author} {\bibfnamefont {M.~W.}\ \bibnamefont
  {Mitchell}},\ and\ \bibinfo {author} {\bibfnamefont {E.}~\bibnamefont
  {Witkowska}},\ }\bibfield  {title} {\bibinfo {title} {Criticality-enhanced
  quantum sensing in ferromagnetic bose-einstein condensates: Role of readout
  measurement and detection noise},\ }\href
  {https://doi.org/10.1103/PhysRevA.103.023317} {\bibfield  {journal} {\bibinfo
   {journal} {Phys. Rev. A}\ }\textbf {\bibinfo {volume} {103}},\ \bibinfo
  {pages} {023317} (\bibinfo {year} {2021})}\BibitemShut {NoStop}%
\bibitem [{\citenamefont {Ilias}\ \emph {et~al.}(2022)\citenamefont {Ilias},
  \citenamefont {Yang}, \citenamefont {Huelga},\ and\ \citenamefont
  {Plenio}}]{2022PRXQuantumcontinus}%
  \BibitemOpen
  \bibfield  {author} {\bibinfo {author} {\bibfnamefont {T.}~\bibnamefont
  {Ilias}}, \bibinfo {author} {\bibfnamefont {D.}~\bibnamefont {Yang}},
  \bibinfo {author} {\bibfnamefont {S.~F.}\ \bibnamefont {Huelga}},\ and\
  \bibinfo {author} {\bibfnamefont {M.~B.}\ \bibnamefont {Plenio}},\ }\bibfield
   {title} {\bibinfo {title} {Criticality-enhanced quantum sensing via
  continuous measurement},\ }\href
  {https://doi.org/10.1103/PRXQuantum.3.010354} {\bibfield  {journal} {\bibinfo
   {journal} {PRX Quantum}\ }\textbf {\bibinfo {volume} {3}},\ \bibinfo {pages}
  {010354} (\bibinfo {year} {2022})}\BibitemShut {NoStop}%
\bibitem [{\citenamefont {Garbe}\ \emph
  {et~al.}(2022{\natexlab{a}})\citenamefont {Garbe}, \citenamefont {Abah},
  \citenamefont {Felicetti},\ and\ \citenamefont
  {Puebla}}]{2022_Garbe_heisenbegkible}%
  \BibitemOpen
  \bibfield  {author} {\bibinfo {author} {\bibfnamefont {L.}~\bibnamefont
  {Garbe}}, \bibinfo {author} {\bibfnamefont {O.}~\bibnamefont {Abah}},
  \bibinfo {author} {\bibfnamefont {S.}~\bibnamefont {Felicetti}},\ and\
  \bibinfo {author} {\bibfnamefont {R.}~\bibnamefont {Puebla}},\ }\bibfield
  {title} {\bibinfo {title} {Critical quantum metrology with fully-connected
  models: from heisenberg to kibble–zurek scaling},\ }\href
  {https://doi.org/10.1088/2058-9565/ac6ca5} {\bibfield  {journal} {\bibinfo
  {journal} {Quantum Science and Technology}\ }\textbf {\bibinfo {volume}
  {7}},\ \bibinfo {pages} {035010} (\bibinfo {year}
  {2022}{\natexlab{a}})}\BibitemShut {NoStop}%
\bibitem [{\citenamefont {Ying}\ \emph {et~al.}(2022)\citenamefont {Ying},
  \citenamefont {Felicetti}, \citenamefont {Liu},\ and\ \citenamefont
  {Braak}}]{2022entropycritical}%
  \BibitemOpen
  \bibfield  {author} {\bibinfo {author} {\bibfnamefont {Z.-J.}\ \bibnamefont
  {Ying}}, \bibinfo {author} {\bibfnamefont {S.}~\bibnamefont {Felicetti}},
  \bibinfo {author} {\bibfnamefont {G.}~\bibnamefont {Liu}},\ and\ \bibinfo
  {author} {\bibfnamefont {D.}~\bibnamefont {Braak}},\ }\bibfield  {title}
  {\bibinfo {title} {Critical quantum metrology in the non-linear quantum rabi
  model},\ }\bibfield  {journal} {\bibinfo  {journal} {Entropy}\ }\textbf
  {\bibinfo {volume} {24}},\ \href {https://doi.org/10.3390/e24081015}
  {10.3390/e24081015} (\bibinfo {year} {2022})\BibitemShut {NoStop}%
\bibitem [{\citenamefont {Ding}\ \emph {et~al.}(2022)\citenamefont {Ding},
  \citenamefont {Liu}, \citenamefont {Shi}, \citenamefont {Guo}, \citenamefont
  {M{\o}lmer},\ and\ \citenamefont {Adams}}]{2022dingnature}%
  \BibitemOpen
  \bibfield  {author} {\bibinfo {author} {\bibfnamefont {D.-S.}\ \bibnamefont
  {Ding}}, \bibinfo {author} {\bibfnamefont {Z.-K.}\ \bibnamefont {Liu}},
  \bibinfo {author} {\bibfnamefont {B.-S.}\ \bibnamefont {Shi}}, \bibinfo
  {author} {\bibfnamefont {G.-C.}\ \bibnamefont {Guo}}, \bibinfo {author}
  {\bibfnamefont {K.}~\bibnamefont {M{\o}lmer}},\ and\ \bibinfo {author}
  {\bibfnamefont {C.~S.}\ \bibnamefont {Adams}},\ }\bibfield  {title} {\bibinfo
  {title} {Enhanced metrology at the critical point of a many-body rydberg
  atomic system},\ }\bibfield  {journal} {\bibinfo  {journal} {Nature Physics}\
  }\href {https://doi.org/10.1038/s41567-022-01777-8}
  {10.1038/s41567-022-01777-8} (\bibinfo {year} {2022})\BibitemShut {NoStop}%
\bibitem [{\citenamefont {Aybar}\ \emph {et~al.}(2022)\citenamefont {Aybar},
  \citenamefont {Niezgoda}, \citenamefont {Mirkhalaf}, \citenamefont
  {Mitchell}, \citenamefont {Benedicto~Orenes},\ and\ \citenamefont
  {Witkowska}}]{Aybar2022criticalquantum}%
  \BibitemOpen
  \bibfield  {author} {\bibinfo {author} {\bibfnamefont {E.}~\bibnamefont
  {Aybar}}, \bibinfo {author} {\bibfnamefont {A.}~\bibnamefont {Niezgoda}},
  \bibinfo {author} {\bibfnamefont {S.~S.}\ \bibnamefont {Mirkhalaf}}, \bibinfo
  {author} {\bibfnamefont {M.~W.}\ \bibnamefont {Mitchell}}, \bibinfo {author}
  {\bibfnamefont {D.}~\bibnamefont {Benedicto~Orenes}},\ and\ \bibinfo {author}
  {\bibfnamefont {E.}~\bibnamefont {Witkowska}},\ }\bibfield  {title} {\bibinfo
  {title} {Critical quantum thermometry and its feasibility in spin systems},\
  }\href {https://doi.org/10.22331/q-2022-09-19-808} {\bibfield  {journal}
  {\bibinfo  {journal} {{Quantum}}\ }\textbf {\bibinfo {volume} {6}},\ \bibinfo
  {pages} {808} (\bibinfo {year} {2022})}\BibitemShut {NoStop}%
\bibitem [{\citenamefont {Garbe}\ \emph
  {et~al.}(2022{\natexlab{b}})\citenamefont {Garbe}, \citenamefont {Abah},
  \citenamefont {Felicetti},\ and\ \citenamefont
  {Puebla}}]{2022garbeexponetialsqueezing}%
  \BibitemOpen
  \bibfield  {author} {\bibinfo {author} {\bibfnamefont {L.}~\bibnamefont
  {Garbe}}, \bibinfo {author} {\bibfnamefont {O.}~\bibnamefont {Abah}},
  \bibinfo {author} {\bibfnamefont {S.}~\bibnamefont {Felicetti}},\ and\
  \bibinfo {author} {\bibfnamefont {R.}~\bibnamefont {Puebla}},\ }\bibfield
  {title} {\bibinfo {title} {Exponential time-scaling of estimation precision
  by reaching a quantum critical point},\ }\href
  {https://doi.org/10.1103/PhysRevResearch.4.043061} {\bibfield  {journal}
  {\bibinfo  {journal} {Phys. Rev. Research}\ }\textbf {\bibinfo {volume}
  {4}},\ \bibinfo {pages} {043061} (\bibinfo {year}
  {2022}{\natexlab{b}})}\BibitemShut {NoStop}%
\bibitem [{\citenamefont {Chin}\ \emph {et~al.}(2010)\citenamefont {Chin},
  \citenamefont {Grimm}, \citenamefont {Julienne},\ and\ \citenamefont
  {Tiesinga}}]{feschbach2010rmp}%
  \BibitemOpen
  \bibfield  {author} {\bibinfo {author} {\bibfnamefont {C.}~\bibnamefont
  {Chin}}, \bibinfo {author} {\bibfnamefont {R.}~\bibnamefont {Grimm}},
  \bibinfo {author} {\bibfnamefont {P.}~\bibnamefont {Julienne}},\ and\
  \bibinfo {author} {\bibfnamefont {E.}~\bibnamefont {Tiesinga}},\ }\bibfield
  {title} {\bibinfo {title} {Feshbach resonances in ultracold gases},\ }\href
  {https://doi.org/10.1103/RevModPhys.82.1225} {\bibfield  {journal} {\bibinfo
  {journal} {Rev. Mod. Phys.}\ }\textbf {\bibinfo {volume} {82}},\ \bibinfo
  {pages} {1225} (\bibinfo {year} {2010})}\BibitemShut {NoStop}%
\bibitem [{\citenamefont {Koch}\ \emph {et~al.}(2022)\citenamefont {Koch},
  \citenamefont {Menon}, \citenamefont {Cuestas}, \citenamefont {Barbosa},
  \citenamefont {Lutz}, \citenamefont {Fogarty}, \citenamefont {Busch},\ and\
  \citenamefont {Widera}}]{Menon2022bcsbec}%
  \BibitemOpen
  \bibfield  {author} {\bibinfo {author} {\bibfnamefont {J.}~\bibnamefont
  {Koch}}, \bibinfo {author} {\bibfnamefont {K.}~\bibnamefont {Menon}},
  \bibinfo {author} {\bibfnamefont {E.}~\bibnamefont {Cuestas}}, \bibinfo
  {author} {\bibfnamefont {S.}~\bibnamefont {Barbosa}}, \bibinfo {author}
  {\bibfnamefont {E.}~\bibnamefont {Lutz}}, \bibinfo {author} {\bibfnamefont
  {T.}~\bibnamefont {Fogarty}}, \bibinfo {author} {\bibfnamefont
  {T.}~\bibnamefont {Busch}},\ and\ \bibinfo {author} {\bibfnamefont
  {A.}~\bibnamefont {Widera}},\ }\href
  {https://doi.org/10.48550/ARXIV.2209.14202} {\bibinfo {title} {Making
  statistics work: a quantum engine in the bec-bcs crossover}} (\bibinfo {year}
  {2022})\BibitemShut {NoStop}%
\bibitem [{\citenamefont {Hern\'andez~Yanes}\ \emph {et~al.}(2022)\citenamefont
  {Hern\'andez~Yanes}, \citenamefont {P\l{}odzie\ifmmode~\acute{n}\else
  \'{n}\fi{}}, \citenamefont {Mackoit Sinkevi\ifmmode \check{c}\else
  \v{c}\fi{}ien\ifmmode~\dot{e}\else \.{e}\fi{}}, \citenamefont
  {\ifmmode~\check{Z}\else \v{Z}\fi{}labys}, \citenamefont
  {Juzeli\ifmmode~\bar{u}\else \={u}\fi{}nas},\ and\ \citenamefont
  {Witkowska}}]{2022twoaxisemilia}%
  \BibitemOpen
  \bibfield  {author} {\bibinfo {author} {\bibfnamefont {T.}~\bibnamefont
  {Hern\'andez~Yanes}}, \bibinfo {author} {\bibfnamefont {M.}~\bibnamefont
  {P\l{}odzie\ifmmode~\acute{n}\else \'{n}\fi{}}}, \bibinfo {author}
  {\bibfnamefont {M.}~\bibnamefont {Mackoit Sinkevi\ifmmode \check{c}\else
  \v{c}\fi{}ien\ifmmode~\dot{e}\else \.{e}\fi{}}}, \bibinfo {author}
  {\bibfnamefont {G.}~\bibnamefont {\ifmmode~\check{Z}\else \v{Z}\fi{}labys}},
  \bibinfo {author} {\bibfnamefont {G.}~\bibnamefont
  {Juzeli\ifmmode~\bar{u}\else \={u}\fi{}nas}},\ and\ \bibinfo {author}
  {\bibfnamefont {E.}~\bibnamefont {Witkowska}},\ }\bibfield  {title} {\bibinfo
  {title} {One- and two-axis squeezing via laser coupling in an atomic
  fermi-hubbard model},\ }\href
  {https://doi.org/10.1103/PhysRevLett.129.090403} {\bibfield  {journal}
  {\bibinfo  {journal} {Phys. Rev. Lett.}\ }\textbf {\bibinfo {volume} {129}},\
  \bibinfo {pages} {090403} (\bibinfo {year} {2022})}\BibitemShut {NoStop}%
\bibitem [{\citenamefont {Napolitano}\ \emph {et~al.}(2011)\citenamefont
  {Napolitano}, \citenamefont {Koschorreck}, \citenamefont {Dubost},
  \citenamefont {Behbood}, \citenamefont {Sewell},\ and\ \citenamefont
  {Mitchell}}]{2011INTbeyondHL}%
  \BibitemOpen
  \bibfield  {author} {\bibinfo {author} {\bibfnamefont {M.}~\bibnamefont
  {Napolitano}}, \bibinfo {author} {\bibfnamefont {M.}~\bibnamefont
  {Koschorreck}}, \bibinfo {author} {\bibfnamefont {B.}~\bibnamefont {Dubost}},
  \bibinfo {author} {\bibfnamefont {N.}~\bibnamefont {Behbood}}, \bibinfo
  {author} {\bibfnamefont {R.~J.}\ \bibnamefont {Sewell}},\ and\ \bibinfo
  {author} {\bibfnamefont {M.~W.}\ \bibnamefont {Mitchell}},\ }\bibfield
  {title} {\bibinfo {title} {Interaction-based quantum metrology showing
  scaling beyond the heisenberg limit},\ }\href
  {https://doi.org/10.1038/nature09778} {\bibfield  {journal} {\bibinfo
  {journal} {Nature}\ }\textbf {\bibinfo {volume} {471}},\ \bibinfo {pages}
  {486} (\bibinfo {year} {2011})}\BibitemShut {NoStop}%
\bibitem [{\citenamefont {Soderberg}\ and\ \citenamefont
  {Monroe}(2010)}]{Soderberg_2010_review}%
  \BibitemOpen
  \bibfield  {author} {\bibinfo {author} {\bibfnamefont {K.-A.~B.}\
  \bibnamefont {Soderberg}}\ and\ \bibinfo {author} {\bibfnamefont
  {C.}~\bibnamefont {Monroe}},\ }\bibfield  {title} {\bibinfo {title}
  {Phonon-mediated entanglement for trapped ion quantum computing},\ }\href
  {https://doi.org/10.1088/0034-4885/73/3/036401} {\bibfield  {journal}
  {\bibinfo  {journal} {Reports on Progress in Physics}\ }\textbf {\bibinfo
  {volume} {73}},\ \bibinfo {pages} {036401} (\bibinfo {year}
  {2010})}\BibitemShut {NoStop}%
\bibitem [{\citenamefont {Gietka}\ \emph {et~al.}(2022)\citenamefont {Gietka},
  \citenamefont {Ruks},\ and\ \citenamefont {Busch}}]{Gietka2022understanding}%
  \BibitemOpen
  \bibfield  {author} {\bibinfo {author} {\bibfnamefont {K.}~\bibnamefont
  {Gietka}}, \bibinfo {author} {\bibfnamefont {L.}~\bibnamefont {Ruks}},\ and\
  \bibinfo {author} {\bibfnamefont {T.}~\bibnamefont {Busch}},\ }\bibfield
  {title} {\bibinfo {title} {Understanding and {I}mproving {C}ritical
  {M}etrology. {Q}uenching {S}uperradiant {L}ight-{M}atter {S}ystems {B}eyond
  the {C}ritical {P}oint},\ }\href {https://doi.org/10.22331/q-2022-04-27-700}
  {\bibfield  {journal} {\bibinfo  {journal} {{Quantum}}\ }\textbf {\bibinfo
  {volume} {6}},\ \bibinfo {pages} {700} (\bibinfo {year} {2022})}\BibitemShut
  {NoStop}%
\bibitem [{\citenamefont {Kustura}\ \emph {et~al.}(2022)\citenamefont
  {Kustura}, \citenamefont {Gonzalez-Ballestero}, \citenamefont {Sommer},
  \citenamefont {Meyer}, \citenamefont {Quidant},\ and\ \citenamefont
  {Romero-Isart}}]{2022mechanicalsqueezing}%
  \BibitemOpen
  \bibfield  {author} {\bibinfo {author} {\bibfnamefont {K.}~\bibnamefont
  {Kustura}}, \bibinfo {author} {\bibfnamefont {C.}~\bibnamefont
  {Gonzalez-Ballestero}}, \bibinfo {author} {\bibfnamefont {A.~d. l.~R.}\
  \bibnamefont {Sommer}}, \bibinfo {author} {\bibfnamefont {N.}~\bibnamefont
  {Meyer}}, \bibinfo {author} {\bibfnamefont {R.}~\bibnamefont {Quidant}},\
  and\ \bibinfo {author} {\bibfnamefont {O.}~\bibnamefont {Romero-Isart}},\
  }\bibfield  {title} {\bibinfo {title} {Mechanical squeezing via unstable
  dynamics in a microcavity},\ }\href
  {https://doi.org/10.1103/PhysRevLett.128.143601} {\bibfield  {journal}
  {\bibinfo  {journal} {Phys. Rev. Lett.}\ }\textbf {\bibinfo {volume} {128}},\
  \bibinfo {pages} {143601} (\bibinfo {year} {2022})}\BibitemShut {NoStop}%
\bibitem [{\citenamefont {Gietka}\ and\ \citenamefont
  {Busch}(2021)}]{2021Inverted_Oscillator_Gietka}%
  \BibitemOpen
  \bibfield  {author} {\bibinfo {author} {\bibfnamefont {K.}~\bibnamefont
  {Gietka}}\ and\ \bibinfo {author} {\bibfnamefont {T.}~\bibnamefont {Busch}},\
  }\bibfield  {title} {\bibinfo {title} {Inverted harmonic oscillator dynamics
  of the nonequilibrium phase transition in the dicke model},\ }\href
  {https://doi.org/10.1103/PhysRevE.104.034132} {\bibfield  {journal} {\bibinfo
   {journal} {Phys. Rev. E}\ }\textbf {\bibinfo {volume} {104}},\ \bibinfo
  {pages} {034132} (\bibinfo {year} {2021})}\BibitemShut {NoStop}%
\bibitem [{\citenamefont {Weiss}\ \emph {et~al.}(2021)\citenamefont {Weiss},
  \citenamefont {Roda-Llordes}, \citenamefont {Torrontegui}, \citenamefont
  {Aspelmeyer},\ and\ \citenamefont {Romero-Isart}}]{weiss2021large}%
  \BibitemOpen
  \bibfield  {author} {\bibinfo {author} {\bibfnamefont {T.}~\bibnamefont
  {Weiss}}, \bibinfo {author} {\bibfnamefont {M.}~\bibnamefont {Roda-Llordes}},
  \bibinfo {author} {\bibfnamefont {E.}~\bibnamefont {Torrontegui}}, \bibinfo
  {author} {\bibfnamefont {M.}~\bibnamefont {Aspelmeyer}},\ and\ \bibinfo
  {author} {\bibfnamefont {O.}~\bibnamefont {Romero-Isart}},\ }\bibfield
  {title} {\bibinfo {title} {Large quantum delocalization of a levitated
  nanoparticle using optimal control: applications for force sensing and
  entangling via weak forces},\ }\href@noop {} {\bibfield  {journal} {\bibinfo
  {journal} {Physical Review Letters}\ }\textbf {\bibinfo {volume} {127}},\
  \bibinfo {pages} {023601} (\bibinfo {year} {2021})}\BibitemShut {NoStop}%
\bibitem [{\citenamefont {Ritsch}\ \emph {et~al.}(2013)\citenamefont {Ritsch},
  \citenamefont {Domokos}, \citenamefont {Brennecke},\ and\ \citenamefont
  {Esslinger}}]{2013helmutrmp}%
  \BibitemOpen
  \bibfield  {author} {\bibinfo {author} {\bibfnamefont {H.}~\bibnamefont
  {Ritsch}}, \bibinfo {author} {\bibfnamefont {P.}~\bibnamefont {Domokos}},
  \bibinfo {author} {\bibfnamefont {F.}~\bibnamefont {Brennecke}},\ and\
  \bibinfo {author} {\bibfnamefont {T.}~\bibnamefont {Esslinger}},\ }\bibfield
  {title} {\bibinfo {title} {Cold atoms in cavity-generated dynamical optical
  potentials},\ }\href {https://doi.org/10.1103/RevModPhys.85.553} {\bibfield
  {journal} {\bibinfo  {journal} {Rev. Mod. Phys.}\ }\textbf {\bibinfo {volume}
  {85}},\ \bibinfo {pages} {553} (\bibinfo {year} {2013})}\BibitemShut
  {NoStop}%
\bibitem [{\citenamefont {Gietka}\ \emph {et~al.}(2019)\citenamefont {Gietka},
  \citenamefont {Mivehvar},\ and\ \citenamefont {Ritsch}}]{2019gietkagrav}%
  \BibitemOpen
  \bibfield  {author} {\bibinfo {author} {\bibfnamefont {K.}~\bibnamefont
  {Gietka}}, \bibinfo {author} {\bibfnamefont {F.}~\bibnamefont {Mivehvar}},\
  and\ \bibinfo {author} {\bibfnamefont {H.}~\bibnamefont {Ritsch}},\
  }\bibfield  {title} {\bibinfo {title} {Supersolid-based gravimeter in a ring
  cavity},\ }\href {https://doi.org/10.1103/PhysRevLett.122.190801} {\bibfield
  {journal} {\bibinfo  {journal} {Phys. Rev. Lett.}\ }\textbf {\bibinfo
  {volume} {122}},\ \bibinfo {pages} {190801} (\bibinfo {year}
  {2019})}\BibitemShut {NoStop}%
\bibitem [{\citenamefont {Gietka}\ \emph
  {et~al.}(2021{\natexlab{b}})\citenamefont {Gietka}, \citenamefont
  {Mivehvar},\ and\ \citenamefont {Busch}}]{2021Gietka_njp}%
  \BibitemOpen
  \bibfield  {author} {\bibinfo {author} {\bibfnamefont {K.}~\bibnamefont
  {Gietka}}, \bibinfo {author} {\bibfnamefont {F.}~\bibnamefont {Mivehvar}},\
  and\ \bibinfo {author} {\bibfnamefont {T.}~\bibnamefont {Busch}},\ }\bibfield
   {title} {\bibinfo {title} {Cavity-enhanced magnetometer with a spinor
  bose–einstein condensate},\ }\href
  {https://doi.org/10.1088/1367-2630/abedff} {\bibfield  {journal} {\bibinfo
  {journal} {New Journal of Physics}\ }\textbf {\bibinfo {volume} {23}},\
  \bibinfo {pages} {043020} (\bibinfo {year} {2021}{\natexlab{b}})}\BibitemShut
  {NoStop}%
\bibitem [{\citenamefont {Mivehvar}\ \emph {et~al.}(2021)\citenamefont
  {Mivehvar}, \citenamefont {Piazza}, \citenamefont {Donner},\ and\
  \citenamefont {Ritsch}}]{2021farokhqed}%
  \BibitemOpen
  \bibfield  {author} {\bibinfo {author} {\bibfnamefont {F.}~\bibnamefont
  {Mivehvar}}, \bibinfo {author} {\bibfnamefont {F.}~\bibnamefont {Piazza}},
  \bibinfo {author} {\bibfnamefont {T.}~\bibnamefont {Donner}},\ and\ \bibinfo
  {author} {\bibfnamefont {H.}~\bibnamefont {Ritsch}},\ }\bibfield  {title}
  {\bibinfo {title} {Cavity qed with quantum gases: new paradigms in many-body
  physics},\ }\href {https://doi.org/10.1080/00018732.2021.1969727} {\bibfield
  {journal} {\bibinfo  {journal} {Advances in Physics}\ }\textbf {\bibinfo
  {volume} {70}},\ \bibinfo {pages} {1} (\bibinfo {year} {2021})},\ \Eprint
  {https://arxiv.org/abs/https://doi.org/10.1080/00018732.2021.1969727}
  {https://doi.org/10.1080/00018732.2021.1969727} \BibitemShut {NoStop}%
\bibitem [{\citenamefont {Chu}\ \emph {et~al.}(2021)\citenamefont {Chu},
  \citenamefont {He}, \citenamefont {Thompson},\ and\ \citenamefont
  {Rey}}]{2021amrcavitymetrology}%
  \BibitemOpen
  \bibfield  {author} {\bibinfo {author} {\bibfnamefont {A.}~\bibnamefont
  {Chu}}, \bibinfo {author} {\bibfnamefont {P.}~\bibnamefont {He}}, \bibinfo
  {author} {\bibfnamefont {J.~K.}\ \bibnamefont {Thompson}},\ and\ \bibinfo
  {author} {\bibfnamefont {A.~M.}\ \bibnamefont {Rey}},\ }\bibfield  {title}
  {\bibinfo {title} {Quantum enhanced cavity qed interferometer with partially
  delocalized atoms in lattices},\ }\href
  {https://doi.org/10.1103/PhysRevLett.127.210401} {\bibfield  {journal}
  {\bibinfo  {journal} {Phys. Rev. Lett.}\ }\textbf {\bibinfo {volume} {127}},\
  \bibinfo {pages} {210401} (\bibinfo {year} {2021})}\BibitemShut {NoStop}%
\bibitem [{\citenamefont {Kr\"{a}mer}\ \emph {et~al.}(2018)\citenamefont
  {Kr\"{a}mer}, \citenamefont {Plankensteiner}, \citenamefont {Ostermann},\
  and\ \citenamefont {Ritsch}}]{kramer2018quantumoptics}%
  \BibitemOpen
  \bibfield  {author} {\bibinfo {author} {\bibfnamefont {S.}~\bibnamefont
  {Kr\"{a}mer}}, \bibinfo {author} {\bibfnamefont {D.}~\bibnamefont
  {Plankensteiner}}, \bibinfo {author} {\bibfnamefont {L.}~\bibnamefont
  {Ostermann}},\ and\ \bibinfo {author} {\bibfnamefont {H.}~\bibnamefont
  {Ritsch}},\ }\bibfield  {title} {\bibinfo {title} {{QuantumOptics}.jl: A
  {J}ulia framework for simulating open quantum systems},\ }\href
  {https://doi.org/10.1016/j.cpc.2018.02.004} {\bibfield  {journal} {\bibinfo
  {journal} {Comput. Phys. Commun}\ }\textbf {\bibinfo {volume} {227}},\
  \bibinfo {pages} {109} (\bibinfo {year} {2018})}\BibitemShut {NoStop}%
\end{thebibliography}
\end{document}